\documentclass[notitlepage,a4paper,aps,prd,onecolumn,superscriptaddress,nofootinbib,groupedaddress,longbibliography]{revtex4-1}
\usepackage{amsmath}
\usepackage{amsfonts}
\usepackage{amssymb}
\usepackage{mathtools}
\usepackage{mathrsfs}
\usepackage[utf8]{inputenc}
\usepackage[T1]{fontenc}
\usepackage[colorlinks=true]{hyperref}

\newcommand{\order}[2]{\overset{\mathclap{\scriptscriptstyle #2}}{#1}\vphantom{#1}}
\newcommand{\lc}[1]{\overset{\circ}{#1}\vphantom{#1}}
\newcommand{\dd}{\mathrm{d}}

\begin{document}

\title{Post-Newtonian limit of scalar-torsion theories of gravity as analogue to scalar-curvature theories}

\author{Elena D. Emtsova}
\email{ed.emcova@physics.msu.ru}
\affiliation{Physical Faculty, Lomonosov Moscow State University, Moscow 119991, Russia}
\affiliation{Sternberg Astronomical Institute, Lomonosov Moscow State University, Universitetsky Prospect, 13, Moscow 119991, Russia}

\author{Manuel Hohmann}
\email{manuel.hohmann@ut.ee}
\affiliation{Laboratory of Theoretical Physics, Institute of Physics, University of Tartu, W. Ostwaldi 1, 50411 Tartu, Estonia}

\begin{abstract}
We consider a recently proposed class of extended teleparallel theories of gravity, which entail a scalar field which is non-minimally coupled to the torsion of a flat, metric-compatible connection. This class of scalar-torsion theories of gravity is constructed in analogy to and as a direct extension of the well-studied class of scalar-curvature gravity theories, and has various common features, such as the conformal frame freedom. For this class we determine the parametrized post-Newtonian limit, both for a massive and a massless scalar field. In the massive case, we determine the effective gravitational constant and the post-Newtonian parameter $\gamma$, both of which depend on the distance between the gravitating and test masses. In the massless case, we calculate the full set of parameters and find that only $\gamma$ and $\beta$ potentially deviate from their general relativity values. In particular, we find that for a minimally coupled scalar field the theory becomes indistinguishable from general relativity at this level of the post-Newtonian approximation.
\end{abstract}

\maketitle


\section{Introduction}\label{sec:intro}
One of the most challenging questions in modern gravitational physics is posed by cosmological observations, such as the accelerating expansion of the Universe at present and early times in its history, known as dark energy and inflation, as well as observations of galaxies and the large scale structure, which hint towards the presence of an unknown, dark matter component, which is apparent only by its gravitational effects. Besides models originating from particle physics, a potential explanation of these observations is given by modified gravity theories. An important class of such theories is constituted by scalar-curvature theories of gravity~\cite{Faraoni:2004pi,Fujii:2003pa}. These theories have in common that they contain one or more scalar fields, which in general are non-minimally coupled to the curvature of the Levi-Civita connection arising from the metric geometry of spacetime. The gravitational dynamics of the theory is then determined by the interaction of these fundamental, metric and scalar fields.

A class of such theories of particular interest is defined in terms of four free functions in the action functional, where any specific choice of these functions defines a concrete theory~\cite{Flanagan:2004bz}. A remarkable property of this class of scalar-tensor theories is their behavior under conformal transformations of the metric, which relate different theories within this class to each other. It is an ongoing debate whether these conformally related theories lead to physically equivalent predictions~\cite{Capozziello:2006dj,Catena:2006bd,Faraoni:2006fx,Deruelle:2010ht,Chiba:2013mha,Postma:2014vaa,Faraoni:1998qx,Capozziello:2010sc,Rondeau:2017xck,Bahamonde:2016wmz,Bahamonde:2017kbs,Brooker:2016oqa}. As an important contribution to this debate, a number of invariant quantities have been identified, which can be used to express physical observables independently of the choice of the conformal frame~\cite{Jarv:2014hma,Kuusk:2015dda}.

While the aforementioned class of theories, like many other modified gravity theories, are readily interpreted as modifications of general relativity in its most well-known formulation in terms of the curvature of the torsion-free, metric-compatible Levi-Civita connection, one may consider alternative starting points for modifications. These starting points may be provided by the equivalent formulations of general relativity either in terms of the torsion of a flat, metric-compatible connection, or in terms of the nonmetricity of a flat, torsion-free connection, or even a combination of both~\cite{BeltranJimenez:2019tjy,Jimenez:2019ghw}. These formulations have received increasing interest during the last years due to the fact that they exhibit more similarities to other gauge field theories, thus potentially providing a link to the theories describing the other fundamental interactions of nature. In this article we will focus on so-called teleparallel models of gravity, where torsion takes the role of curvature as the quantity which mediates the gravitational interaction~\cite{Einstein:1928,Moller:1961,Aldrovandi:2013wha,Maluf:2013gaa,Golovnev:2018red}. There are different possibilities how this flat, metric-compatible connection may be implemented. In its original formulation of the teleparallel equivalent of general relativity (TEGR) and its modifications the Weitzenböck connection of a tetrad was assumed, which possesses a vanishing spin connection. While in TEGR the spin connection does not contribute to the field equations, and so there is no harm in a priori fixing it, this it not the case in modified theories. One potential issue arising from this fact is a possible breaking of local Lorentz invariance~\cite{Li:2010cg,Sotiriou:2010mv} and the appearance of spurious degrees of freedom~\cite{Li:2011rn,Ong:2013qja,Izumi:2013dca,Chen:2014qtl}. There are different possibilities to resolve this issue. One such possibility is the use of the Palatini approach to implement the flat, metric-compatible connection~\cite{BeltranJimenez:2018vdo}. Another approach, which is the one we will make use of here, is the covariant formulation, which features an arbitrary, flat, metric-compatible spin connection~\cite{Krssak:2015oua,Golovnev:2017dox,Krssak:2018ywd,Bejarano:2019fii}.

The teleparallel equivalent of general relatively has been the starting point for numerous modified gravity theories, most of which aim at answering the aforementioned questions raised by general relativity in the light of cosmological observations and the tensions with particle physics~\cite{Geng:2011aj,Izumi:2013dca,Chakrabarti:2017moe,Otalora:2013tba,Jamil:2012vb,Chen:2014qsa,Bahamonde:2015hza,Bamba:2013jqa,Nojiri:2017ncd}. While most of these theories, including extensions by scalar fields, make use of the Weitzenböck connection to implement the teleparallel geometry, also a Lorentz covariant formulation of scalar-torsion gravity has recently been proposed~\cite{Hohmann:2018rwf}, and we will make use of this proposal here. It turns out that a large class of scalar-torsion theories can be constructed following this principle of Lorentz covariance~\cite{Hohmann:2018vle,Hohmann:2018dqh,Hohmann:2018ijr}. Focusing on similarities to scalar-curvature theories, one finds a particular subclass of scalar-torsion theories whose action is characterized by one more free function of the scalar field compared to the classical scalar-curvature theories~\cite{Flanagan:2004bz}, and reduces to the latter for a particular choice of this function~\cite{Hohmann:2018ijr}. This class of theories also exhibits invariance under conformal frame rescalings, again in analogy to scalar-curvature gravity. This is the class of theories we will study in this article.

An important criterion allowing to restrict the large class of scalar-torsion theories is their compatibility with observations on smaller scales, such as the solar system. A well established tool for testing the viability of metric theories of gravity is the parameterized post-Newtonian (PPN) formalism~\cite{Will:1993ns,Will:2014kxa,Will:2018bme}, which characterizes any given theory of gravity by a set of ten (usually constant) parameters. Comparing these parameters obtained from a theory with high-precision measurements of their values in solar system experiments thus yields bounds on the allowed classes of theories.

The application of the PPN formalism to particular scalar-torsion gravity theories, in particular to the original teleparallel dark energy model~\cite{Li:2013oef} and theories with general coupling function and potential~\cite{Chen:2014qsa}, has shown that these theories yield the same values for the PPN parameters as general relativity, and thus cannot be distinguished by the aforementioned measurements. However, it has also been found that more general theories, including a non-minimal coupling to the teleparallel boundary term, lead to a deviating post-Newtonian limit~\cite{Sadjadi:2016kwj}. The aim of this article is to extend these earlier studies to the general class of scalar-torsion theories of gravity mentioned above~\cite{Hohmann:2018ijr}. Since we are using the covariant formulation of these theories, we make use of a recently developed adaptation of the PPN formalism to covariant teleparallel gravity theories~\cite{Ualikhanova:2019ygl}, which we further adapt to theories based on a scalar field and a tetrad~\cite{Hayward:1981bk}, thereby building upon numerous previous studies of the post-Newtonian limit of translation and Poincaré gauge theory~\cite{Schweizer:1979up,Schweizer:1980vn,Smalley:1980em,Nitsch:1979qn,Gladchenko:1990nw,Gladchenko:1994wu}. In particular, we study how the coupling to the teleparallel boundary term, which can equivalently be described by a kinetic coupling between the scalar field and the vector torsion, is related to the deviation of the PPN parameters from their general relativity values. During this work we make use of the conformal frame freedom of scalar-torsion gravity to calculate the PPN parameters in the Jordan frame and to express them in terms of frame independent quantities.

The outline of this article is as follows. We start with a brief review of the dynamical variables and field equations of the class of scalar-torsion theories we consider in section~\ref{sec:dynamics}. Another brief review of the PPN formalism is presented in section~\ref{sec:ppn}, together with its adaptation to scalar-torsion gravity. We then come to the main part of the paper, with the derivation of the PPN parameter \(\gamma\) for the massive scalar field case in section~\ref{sec:massive}, as well as the full set of PPN parameters for a massless scalar field in section~\ref{sec:massless}. We apply our results to a few example theories in section~\ref{sec:examples}, before we conclude with a discussion and outlook in section~\ref{sec:conclusion}.

In this article we use uppercase Latin letters \(A, B, \ldots = 0, \ldots, 3\) for Lorentz indices, lowercase Greek letters \(\mu, \nu, \ldots = 0, \ldots, 3\) for spacetime indices and lowercase Latin letters \(i, j, \ldots = 1, \ldots, 3\) for spatial indices. In our convention the Minkowski metric \(\eta_{AB}\) and \(\eta_{\mu\nu}\) has signature \((-,+,+,+)\).

\section{Field variables and their dynamics}\label{sec:dynamics}
We start our discussion of the post-Newtonian limit of a recently proposed class of scalar-torsion theories of gravity~\cite{Hohmann:2018ijr} with a brief review of their field content, action and field equations. These theories make use of the covariant formulation of scalar-torsion gravity~\cite{Hohmann:2018rwf}, where the dynamical fields are given by a tetrad \(\theta^A{}_{\mu}\), a flat Lorentz spin connection \(\omega^A{}_{B\mu}\) and a scalar field \(\phi\). From these field variables one derives the metric
\begin{equation}
g_{\mu\nu} = \eta_{AB}\theta^A{}_{\mu}\theta^B{}_{\nu}
\end{equation}
and the torsion
\begin{equation}
T^{\rho}{}_{\mu\nu} = e_A{}^{\rho}\left(\partial_{\mu}\theta^A{}_{\nu} - \partial_{\nu}\theta^A{}_{\mu} + \omega^A{}_{B\mu}\theta^B{}_{\nu} - \omega^A{}_{B\nu}\theta^B{}_{\mu}\right)\,,
\end{equation}
where \(e_A{}^{\mu}\) is the inverse tetrad defined such that \(\theta^A{}_{\mu}e_A{}^{\nu} = \delta_{\mu}^{\nu}\) and \(\theta^A{}_{\mu}e_B{}^{\mu} = \delta^A_B\). The metric further defines a Levi-Civita connection \(\lc{\nabla}\) together with its respective curvature tensors; all quantities derived from this connection will be denoted with an empty circle.

The action we consider here will be of the form
\begin{equation}\label{eqn:action}
S[\theta^A{}_{\mu}, \omega^A{}_{B\mu}, \phi, \chi^I] = S_g[\theta^A{}_{\mu}, \omega^A{}_{B\mu}, \phi] + S_m[e^{\alpha(\phi)}\theta^A{}_{\mu}, \chi^I]\,,
\end{equation}
and thus splits into a gravitational part \(S_g\) and a matter part \(S_m\). The latter depends, in addition to the aforementioned dynamical fields, on an arbitrary set \(\chi^I\) of matter fields. In this article we will assume the matter source to be given by a perfect fluid, as discussed in detail in the following section~\ref{sec:ppn}. We further assume that there is no direct coupling between the matter fields \(\chi^I\) and the teleparallel spin connection \(\omega^A{}_{B\mu}\), and that the coupling to the tetrad and the scalar field is mediated only via the conformally rescaled metric \(e^{2\alpha(\phi)}g_{\mu\nu} = e^{2\alpha(\phi)}\eta_{AB}\theta^A{}_{\mu}\theta^B{}_{\nu}\) with a free function \(\alpha\) of the scalar field. It follows from this assumption that the variation of the matter action with respect to the dynamical fields, after performing integration by parts, is of the form
\begin{equation}\label{eqn:matactvar}
\delta S_m[e^{\alpha(\phi)}\theta^A{}_{\mu}, \chi^I] = \int_M\left[\Theta_A{}^{\mu}\left(\delta\theta^A{}_{\mu} + \alpha'(\phi)\theta^A{}_{\mu}\delta\phi\right) + \varpi_I\delta\chi^I\right]\theta\,\dd^4x\,,
\end{equation}
where the prime in \(\alpha'(\phi) = \dd\alpha/\dd\phi\) denotes the derivative with respect to \(\phi\), and that the energy-momentum tensor \(\Theta_{\mu\nu} = \theta^A{}_{\mu}g_{\nu\rho}\Theta_A{}^{\rho}\) is symmetric, \(\Theta_{[\mu\nu]} = 0\). Here \(\theta\) denotes the determinant of the tetrad \(\theta^A{}_{\mu}\), and \(\varpi_I = 0\) are the matter field equations.

For the gravitational part of the action we assume the form
\begin{equation}\label{eqn:actiong}
S_g[\theta^A{}_{\mu}, \omega^A{}_{B\mu}, \phi] = \frac{1}{2\kappa^2}\int_M\left[-\mathcal{A}(\phi)T + 2\mathcal{B}(\phi)X + 2\mathcal{C}(\phi)Y - 2\kappa^2\mathcal{V}(\phi)\right]\theta\,\dd^4x\,,
\end{equation}
where the torsion scalar
\begin{equation}\label{eqn:torsscal}
T = \frac{1}{2}T^{\rho}{}_{\mu\nu}S_{\rho}{}^{\mu\nu}
\end{equation}
is defined via the superpotential
\begin{equation}\label{eqn:suppot}
S_{\rho\mu\nu} = \frac{1}{2}\left(T_{\nu\mu\rho} + T_{\rho\mu\nu} - T_{\mu\nu\rho}\right) - g_{\rho\mu}T^{\sigma}{}_{\sigma\nu} + g_{\rho\nu}T^{\sigma}{}_{\sigma\mu}\,,
\end{equation}
and we have used the scalar field kinetic term
\begin{equation}\label{eqn:defx}
X = -\frac{1}{2}g^{\mu\nu}\phi_{,\mu}\phi_{,\nu}\,,
\end{equation}
as well as the derivative coupling term
\begin{equation}\label{eqn:defy}
Y = g^{\mu\nu}T^{\rho}{}_{\rho\mu}\phi_{,\nu}\,.
\end{equation}
Any particular action of this class is defined by a choice of the free functions \(\mathcal{A}, \mathcal{B}, \mathcal{C}, \mathcal{V}\) of the scalar field, in addition to the free function \(\alpha\) in the matter action. The combined action~\eqref{eqn:action} keeps its form under conformal transformations \(\bar{\theta}^A{}_{\mu} = e^{\gamma(\phi)}\theta^A{}_{\mu}\) of the tetrad and redefinitions \(\bar{\phi} = f(\phi)\) of the scalar field~\cite{Hohmann:2018ijr}, with arbitrary functions \(\gamma(\phi)\) and \(f(\phi)\). This allows us to reduce the number of free functions in the action. During the remainder of this article we will choose to work in the Jordan frame, and hence assume \(\alpha(\phi) \equiv 0\), which can be achieved by performing a conformal transformation with \(\gamma = \alpha\).

We also remark that the derivative coupling term, which is necessary to render the gravitational action invariant under conformal transformations as shown in~\cite{Hohmann:2018ijr}, can also be written as
\begin{equation}\label{eqn:derivbound}
2\mathcal{C}(\phi)Y = 2\lc{\nabla}_{\nu}[\tilde{\mathcal{C}}(\phi)T_{\mu}{}^{\mu\nu}] - \tilde{\mathcal{C}}(\phi)B\,,
\end{equation}
where \(\mathcal{C} = \tilde{\mathcal{C}}'\) and we introduced the teleparallel boundary term
\begin{equation}\label{eqn:boundary}
B = \lc{R} + T = 2\lc{\nabla}_{\nu}T_{\mu}{}^{\mu\nu}\,.
\end{equation}
Omitting the total divergence in~\eqref{eqn:derivbound}, we may thus regard the derivative coupling as a coupling to the boundary term \(B\) instead.

Following the derivation detailed in~\cite{Hohmann:2018ijr}, we can decompose the field equations into three separate sets of equations, which we write in the form
\begin{equation}\label{eqn:fieldeqns}
E_{(\mu\nu)} = \kappa^2\Theta_{\mu\nu}\,, \quad
E_{[\mu\nu]} = 0\,, \quad
E = 0\,.
\end{equation}
The first two equations are obtained by variation of the action with respect to the tetrad \(\theta^A{}_{\mu}\), further using the tetrad and the metric in order to convert both indices to lower spacetime indices, and finally splitting the resulting equations into their symmetric and antisymmetric parts. Their left hand side, obtained from variation of the gravitational part~\eqref{eqn:actiong} of the action, reads
\begin{multline}\label{eqn:clafeqtets}
E_{(\mu\nu)} = \left(\mathcal{A}' + \mathcal{C}\right)S_{(\mu\nu)}{}^{\rho}\phi_{,\rho} + \mathcal{A}\left(\lc{R}_{\mu\nu} - \frac{1}{2}\lc{R}g_{\mu\nu}\right) + \left(\frac{1}{2}\mathcal{B} - \mathcal{C}'\right)\phi_{,\rho}\phi_{,\sigma}g^{\rho\sigma}g_{\mu\nu}\\
- (\mathcal{B} - \mathcal{C}')\phi_{,\mu}\phi_{,\nu} + \mathcal{C}\left(\lc{\nabla}_{\mu}\lc{\nabla}_{\nu}\phi - \lc{\square}\phi g_{\mu\nu}\right) + \kappa^2\mathcal{V}g_{\mu\nu}
\end{multline}
for the symmetric part and
\begin{equation}\label{eqn:clafeqcon}
E_{[\mu\nu]} = (\mathcal{A}' + \mathcal{C})T^{\rho}{}_{[\mu\nu}\phi_{,\rho]}
\end{equation}
for the antisymmetric part. The latter field equation can also be obtained by varying the action with respect to the spin connection \(\omega^A{}_{B\mu}\), allowing only such variations which preserve its vanishing curvature. The third and last field equation is obtained by variation with respect to the scalar field. Its left hand side takes the form
\begin{equation}\label{eqn:clafeqscal}
E = \frac{1}{2}\mathcal{A}'T - \mathcal{B}\lc{\square}\phi - \frac{1}{2}\mathcal{B}'g^{\mu\nu}\phi_{,\mu}\phi_{,\nu} + \mathcal{C}\lc{\nabla}_{\mu}T_{\nu}{}^{\nu\mu} + \kappa^2\mathcal{V}'\,.
\end{equation}
While it would be possible to solve these equations directly using the perturbative expansion discussed in the following section, it turns out that one can significantly simplify this task by performing a number of transformations on the field equations. We start by replacing the symmetric part of the field equations by its trace-reversed form, which we define as
\begin{equation}\label{eqn:trfieldeqns}
\bar{E}_{(\mu\nu)} = \kappa^2\bar{\Theta}_{\mu\nu}\,, \quad
\bar{E}_{(\mu\nu)} = E_{(\mu\nu)} - \frac{1}{2}g_{\mu\nu}g^{\rho\sigma}E_{\rho\sigma}\,, \quad
\bar{\Theta}_{\mu\nu} = \Theta_{\mu\nu} - \frac{1}{2}g_{\mu\nu}g^{\rho\sigma}\Theta_{\rho\sigma}\,.
\end{equation}
After this transformation the left hand side of the corresponding field equations reads
\begin{equation}\label{eqn:clafeqtrv}
\bar{E}_{(\mu\nu)}  =(\mathcal{A}' + \mathcal{C})\left(S_{(\mu \nu)} {}^\rho + g_{\mu \nu }T_{\chi}{}^{\chi \rho}\right)\phi_{,\rho} +\mathcal{A}\lc{R}_{\mu \nu} + \frac{1}{2} \mathcal{C}' g_{\mu \nu} \phi_{,\rho} \phi_{,\sigma} g^{\rho \sigma}
         - (\mathcal{B} -\mathcal{C}') \phi_{,\mu} \phi_{,\nu}  + \mathcal{C} \lc{\nabla}_{\mu}\lc{\nabla}_{\nu} \phi   +\frac{1}{2} \mathcal{C} \lc{\square}\phi g_{\mu \nu} - \kappa^2 \mathcal{V} g_{\mu \nu}
      \,.
\end{equation}
The second transformation we apply concerns the scalar field equation. Note that the left hand side~\eqref{eqn:clafeqscal} contains second order derivatives of the tetrad, which enter through the covariant derivative of the torsion. These can be eliminated by adding a suitable multiple of the trace of the tetrad field equations, so that the transformed scalar field equation takes the form
\begin{equation}\label{eqn:debfieldeqns}
\bar{E} = 2\mathcal{A}E + \mathcal{C}g^{\mu\nu}E_{\mu\nu} = \kappa^2\mathcal{C}\Theta\,,
\end{equation}
where the left hand side is now given by
\begin{equation}\label{eqn:clafeqdeb}
\bar{E} = (\mathcal{A}' + \mathcal{C})\left(\mathcal{A}T - 2\mathcal{C}T_{\mu}{}^{\mu\nu}\phi_{,\nu}\right) - \left(2\mathcal{A}\mathcal{B} + 3\mathcal{C}^2\right)\lc{\square}\phi + (\mathcal{B}\mathcal{C} - \mathcal{A}\mathcal{B}' - 3\mathcal{C}\mathcal{C}')g^{\mu\nu}\phi_{,\mu}\phi_{,\nu} + 2\kappa^2(\mathcal{A}\mathcal{V}' + 2\mathcal{C}\mathcal{V})\,.
\end{equation}
These are the equations we will use in the remainder of this article. In order to solve them, we will perform a perturbative expansion of the dynamical fields. This will be discussed in the following section.

\section{Post-Newtonian approximation}\label{sec:ppn}
We continue with a brief review of the parametrized post-Newtonian formalism~\cite{Will:1993ns,Will:2014kxa,Will:2018bme}, which we will then apply to the class of theories discussed in the previous section. Note that there are different versions of this formalism; here we will use the notation and definitions in its classical form~\cite{Will:1993ns}. This formalism has recently been adapted to the covariant formulation of teleparallel gravity theories~\cite{Ualikhanova:2019ygl}, which will be the basis for the formalism we use here. For our purpose we further adapt the formalism to also include a scalar field besides the tetrad~\cite{Hayward:1981bk}.

Basic assumption of the PPN formalism is that the energy-momentum tensor corresponds to a perfect fluid with rest energy density \(\rho\), specific internal energy \(\Pi\), pressure \(p\) and four-velocity \(u^{\mu}\), which is given by
\begin{equation}\label{eqn:tmunu}
\Theta^{\mu\nu} = (\rho + \rho\Pi + p)u^{\mu}u^{\nu} + pg^{\mu\nu}\,.
\end{equation}
The four-velocity \(u^{\mu}\) is normalized by the metric \(g_{\mu\nu}\), so that \(u^{\mu}u^{\nu}g_{\mu\nu} = -1\). One then assumes that there exists a given frame of reference, conventionally identified with the universe rest frame, in which the velocity \(v^i = u^i/u^0\) of the source matter is small compared to the speed of light, which we set to unity, \(c \equiv 1\). One then introduces velocity orders \(\mathcal{O}(n) \propto |\vec{v}|^n\) as perturbation parameter in which all dynamical quantities are expanded. Here we choose to work in the Weitzenböck gauge, so that \(\omega^A{}_{B\mu} \equiv 0\) at all perturbation orders~\cite{Ualikhanova:2019ygl}, and we must expand only the tetrad and the scalar field. The zeroth order \(\mathcal{O}(0)\) is the background solution, for which we choose the diagonal tetrad \(\Delta^A{}_{\mu} = \mathrm{diag}(1, 1, 1, 1)\), as well as a constant background value \(\Phi\) of the scalar field. We then write the perturbative expansion of the tetrad in the form
\begin{equation}\label{eqn:tetradexp}
\theta^A{}_{\mu} = \Delta^A{}_{\mu} + \tau^A{}_{\mu} = \Delta^A{}_{\mu} + \order{\tau}{1}^A{}_{\mu} + \order{\tau}{2}^A{}_{\mu} + \order{\tau}{3}^A{}_{\mu} + \order{\tau}{4}^A{}_{\mu} + \mathcal{O}(5)\,,
\end{equation}
while the scalar field \(\phi\) expansion reads
\begin{equation}\label{eqn:scalarexp}
\phi = \Phi + \psi = \Phi + \order{\psi}{1} + \order{\psi}{2} + \order{\psi}{3} + \order{\psi}{4} + \mathcal{O}(5)\,.
\end{equation}
Here we have used overscript numbers to denote velocity orders, i.e., each term \(\order{\tau}{n}^A{}_{\mu}\) resp. \(\order{\psi}{n}\) is of order \(\mathcal{O}(n)\). Velocity orders beyond the fourth order are not considered in the PPN formalism and will not be necessary for the derivation of the PPN parameters.

Together with the scalar field we also have to expand the free functions \(\mathcal{A}, \mathcal{B}, \mathcal{C}, \mathcal{V}\) in the action~\eqref{eqn:actiong} into velocity orders. For this purpose we use a Taylor expansion of the form
\begin{equation}\label{eqn:taylorseries}
\mathcal{A}(\phi) = \sum_{k = 0}^{\infty}\frac{\psi^k}{k!}\left.\frac{d^k}{d\phi^k}\mathcal{A}(\phi)\right|_{\phi = \Phi}\,,
\end{equation}
and analogously for the other functions, and introduce roman letters instead of the script letters to denote the values of the derivatives at the cosmological background value of the scalar field, which appear in the Taylor coefficients, i.e.,
\begin{equation}\label{eqn:taylorcoeff}
A = \mathcal{A}(\Phi)\,, \quad
A' = \mathcal{A}'(\Phi)\,, \quad
A'' = \mathcal{A}''(\Phi)\,, \quad
A''' = \mathcal{A}'''(\Phi)\,, \quad \ldots
\end{equation}
The Taylor series~\eqref{eqn:taylorseries} is then further expanded into velocity orders, where all Taylor coefficients are assumed to be of velocity order \(\mathcal{O}(0)\).

Further following the teleparallel PPN formalism~\cite{Ualikhanova:2019ygl}, we introduce another convenient expression for the tetrad perturbations. For this purpose we lower the Lorentz index using the Minkowski metric \(\eta_{AB}\) and convert it into a spacetime index using the background tetrad \(\Delta^A{}_{\mu}\). This yields the expressions
\begin{equation}
\tau_{\mu\nu} = \Delta^A{}_{\mu}\eta_{AB}\tau^B{}_{\nu}\,, \quad
\order{\tau}{n}_{\mu\nu} = \Delta^A{}_{\mu}\eta_{AB}\order{\tau}{n}^B{}_{\nu}\,,
\end{equation}
with pure spacetime indices. In order to determine the PPN parameters, not all components of the tetrad and the scalar field need to be expanded to the fourth velocity order, while others vanish due to Newtonian energy conservation or time reversal symmetry. For the scalar-torsion model, it turns out that the only relevant, non-vanishing components of the field variables we need to determine in this article are given by
\begin{equation}\label{eqn:ppnfields}
\order{\tau}{2}_{00}\,, \quad
\order{\tau}{2}_{ij}\,, \quad
\order{\tau}{3}_{0i}\,, \quad
\order{\tau}{3}_{i0}\,, \quad
\order{\tau}{4}_{00}\,, \quad
\order{\psi}{2}\,, \quad
\order{\psi}{4}\,.
\end{equation}
Using the perturbative expansion~\eqref{eqn:tetradexp} and the tetrad components listed above we can expand all terms appearing in the field equations up to their relevant velocity orders. Of most importance for our calculation is the metric, whose background solution follows from the diagonal background tetrad \(\Delta^A{}_{\mu}\) to be a flat Minkowski metric, \(\order{g}{0}_{\mu\nu} = \eta_{\mu\nu}\), and which is expanded around this background in terms of velocity orders in the form
\begin{equation}\label{eqn:metricexp}
\order{g}{2}_{00} = 2\order{\tau}{2}_{00}\,, \quad
\order{g}{2}_{ij} = 2\order{\tau}{2}_{(ij)}\,, \quad
\order{g}{3}_{0i} = 2\order{\tau}{3}_{(i0)}\,, \quad
\order{g}{4}_{00} = -(\order{\tau}{2}_{00})^2 + 2\order{\tau}{4}_{00}\,.
\end{equation}
In the field equations further appears the torsion, which can be expanded in the form~\cite{Ualikhanova:2019ygl}
\begin{equation}\label{eqn:torsionexp}
\order{T}{2}^0{}_{0i} = \order{\tau}{2}_{00,i}\,, \quad
\order{T}{2}^i{}_{jk} = 2\delta^{il}\order{\tau}{2}_{l[k,j]}\,, \quad
\order{T}{3}^i{}_{0j} = \delta^{ik}(\order{\tau}{2}_{kj,0} - \order{\tau}{3}_{k0,j})\,, \quad
\order{T}{3}^0{}_{ij} = 2\order{\tau}{3}_{0[i,j]}\,, \quad
\order{T}{4}^0{}_{0i} = \order{\tau}{2}_{00}\order{\tau}{2}_{00,i} - \order{\tau}{3}_{0i,0} + \order{\tau}{4}_{00,i}\,.
\end{equation}
For the derivatives of the tetrad and the scalar field we need the additional assumption that the gravitational field is quasi-static, so that changes are only induced by the motion of the source matter. Time derivatives \(\partial_0\) of the tetrad components and scalar field are therefore weighted with an additional velocity order \(\mathcal{O}(1)\).

Finally, we use the expansion~\eqref{eqn:metricexp} of the metric tensor in order to expand the energy-momentum tensor~\eqref{eqn:tmunu} into velocity orders and tetrad perturbations. Using the standard PPN assignment of velocity orders also to the rest mass density, specific internal energy and pressure, which is based on their orders of magnitude in the solar system, and which assigns velocity orders \(\mathcal{O}(2)\) to \(\rho\) and \(\Pi\) and \(\mathcal{O}(4)\) to \(p\)~\cite{Will:1993ns}, the energy-momentum tensor~\eqref{eqn:tmunu} can then be expanded in the form
\begin{subequations}\label{eqn:energymomentum}
\begin{align}
\Theta_{00} &= \rho\left(1 + \Pi + v^2 - 2\order{\tau}{2}_{00}\right) + \mathcal{O}(6)\,,\\
\Theta_{0j} &= -\rho v_j + \mathcal{O}(5)\,,\\
\Theta_{ij} &= \rho v_iv_j + p\delta_{ij} + \mathcal{O}(6)\,.
\end{align}
\end{subequations}
For later use we also expand the trace-reversed energy momentum tensor introduced in the field equations~\eqref{eqn:trfieldeqns} in terms of velocity orders and obtain the expressions
\begin{subequations}\label{eqn:trenergymomentum}
\begin{align}
\bar{\Theta}_{00} &= \frac{1}{2}\rho + \frac{1}{2}\rho\Pi + \rho v^2 - \rho\order{\tau}{2}_{00} + \frac{3}{2}p + \mathcal{O}(6)\,,\\
\bar{\Theta}_{0j} &= -\rho v_j + \mathcal{O}(5)\,,\\
\bar{\Theta}_{ij} &= \frac{1}{2}\rho\delta_{ij} + \frac{1}{2}\rho\Pi\delta_{ij} + \rho v_iv_j + \rho\order{\tau}{2}_{(ij)} - \frac{1}{2}p\delta_{ij} + \mathcal{O}(6)\,.
\end{align}
\end{subequations}
Note in particular that at the zeroth velocity order the energy-momentum tensor vanishes, \(\order{\Theta}{0}_{\mu\nu} = 0\), so that we are left with solving the vacuum field equations. Inserting our assumed background values \(\order{\theta}{0}^A{}_{\mu} = \Delta^A{}_{\mu}\) for the tetrad and \(\order{\phi}{0} = \Phi\) into the respective field equations~\eqref{eqn:fieldeqns}, we find that their gravitational part at the zeroth order is given by
\begin{equation}
\order{E}{0}_{00} = -\kappa^2V\,, \quad
\order{E}{0}_{ij} = \kappa^2V\delta_{ij}\,, \quad
\order{E}{0} = \kappa^2V'\,.
\end{equation}
It thus follows that the perturbation ansatz is consistent with the vacuum field equations only if the parameter functions satisfy \(V = V' = 0\). We will therefore restrict ourselves to theories satisfying these condition during the remaining sections of this article. While these conditions may seem very restrictive at first sight, this is not necessarily the case. The condition \(V = 0\) simply implies that any cosmological constant is sufficiently small to leave the solar system unaffected, which is a reasonable assumption. Further, \(V' = 0\) may appear as an attractor in scalar-torsion cosmology, and is therefore also reasonable in the late universe~\cite{Jarv:2015odu}.

\section{Massive case: PPN parameter $\gamma$}\label{sec:massive}
We now come to the derivation of the post-Newtonian limit of the class of scalar-torsion theories displayed in section~\ref{sec:dynamics}. In this section we will consider a general potential \(\mathcal{V}\) for the scalar field, on which we impose no restrictions except for the consistency conditions \(V = V' = 0\) explained in the preceding section. To solve the perturbative field equations, we consider the simple case of a static point mass as the source matter, which we explain in section~\ref{ssec:mvsource}. Under this assumption we solve the field equations at the second velocity order: for the scalar field in section~\ref{ssec:mvscal2}, and for the time and space components of the tetrad in sections~\ref{ssec:mvtett2} and~\ref{ssec:mvtets2}, respectively. From these solutions we can determine the second order metric perturbations and PPN parameter \(\gamma\) in section~\ref{ssec:mvmetric}.

\subsection{Static point mass source}\label{ssec:mvsource}
The starting point of our calculation is the assumption that the source of the gravitational field is given by a single point-like mass \(M\), whose energy-momentum tensor is of the form~\eqref{eqn:tmunu} with
\begin{equation}\label{eqn:pointmass}
\rho = M\delta(\vec{x})\,, \quad \Pi = 0\,, \quad p = 0\,, \quad v_i = 0\,.
\end{equation}
We thus assume that the point mass is at rest in our chosen coordinate system. In the following, we will use spherical coordinates, with \(r\) denoting the radial coordinate, and the point mass located at the origin \(r = 0\). Further, we will denote by \(U(r) = M/r\) the Newtonian gravitational potential of this source.

\subsection{Scalar field at second order}\label{ssec:mvscal2}
At the second order we can write the equation~\eqref{eqn:debfieldeqns} with~\eqref{eqn:clafeqdeb} for the scalar field in the form
\begin{equation}\label{eqn:psi2mass}
\triangle\order{\psi}{2} -m^2_\phi \order{\psi}{2} = -c_\phi\rho\,,
\end{equation}
where \(\triangle = \delta^{ij}\partial_i\partial_j\) is the Laplace operator and we have introduced the abbreviations
\begin{equation}\label{eqn:scalmass}
m^2_\phi=\frac{2 \kappa^2 A V''}{2 AB +3 C^2}\,, \quad
c_\phi=\frac{-\kappa^2 C}{2 AB +3 C^2}\,.
\end{equation}
We see that the equation is given by a screened Poisson equation, where \(m_{\phi}\) can be interpreted as the mass of the scalar field. The solution of the second order equation~\eqref{eqn:psi2mass} is thus given by
\begin{equation}\label{eqn:psi2solmv}
\order{\psi}{2} (r)= \cfrac{M}{4 \pi r} c_\phi e^{-m_\phi r}\,,
\end{equation}
which has the form of a Yukawa potential.

\subsection{Temporal tetrad components at second order}\label{ssec:mvtett2}
In the next step we can write the second order of the trace-reversed tetrad field equation~\eqref{eqn:trfieldeqns} with~\eqref{eqn:clafeqtrv} for the time component $\order{\bar{E}}{2}_{00}$, which takes the form
\begin{equation}\label{eqn:tau002mass}
2\triangle\order{\tau}{2}_{00} =c_1 \order{\psi}{2} - c_2 \rho\,.
\end{equation}
Here we have introduced the additional abbreviations
\begin{equation}
c_1=-\frac{C}{A} m^2_\phi\,, \quad
c_2=-\frac{c_\phi  C }{A} +\frac{\kappa^2}{A}\,.
\end{equation}
Substituting the previously found solution~\eqref{eqn:psi2solmv} for $\order{\psi}{2}$ into~\eqref{eqn:tau002mass} and then solving it we get
\begin{equation}
\order{\tau}{2}{}_{00} = G_{\text{eff}}(r)U(r) = \cfrac{M}{8 \pi r} \left[c_2+\frac{c_1 c_\phi}{m^2_\phi} (e^{-m_\phi r} -1) \right]\,.
\end{equation}
In the expression above we have introduced the effective gravitational constant, which is given by
\begin{equation}
G_{\text{eff}} (r) = \frac{1}{8 \pi} \left[c_2+\frac{c_1 c_\phi}{m^2_\phi} (e^{-m_\phi r} -1) \right]\,,
\end{equation}
and hence depends on the distance \(r\) between the observer (or test mass) and the gravitating mass. We see that the effective gravitational potential consists of the superposition of a pure Newtonian potential and a Yukawa-type term. Further, we find that in the case \(C \to 0\) no Yukawa term arises, and the gravitational constant becomes truly a constant, \(G_{\text{eff}} \to \kappa^2/8\pi A\). This is related to the fact that in this case the source term in the scalar field equation vanishes, and hence no second-order scalar field is excited. However, also in this case the gravitational constant still depends on the background value \(A\) of the function \(\mathcal{A}\) in front of the torsion scalar \(T\) in the action~\eqref{eqn:actiong}, as one expects~\cite{Jarv:2017npl}.

\subsection{Spatial tetrad components at second order}\label{ssec:mvtets2}
We then come to the second order of the trace-reversed tetrad field equation~\eqref{eqn:trfieldeqns} with~\eqref{eqn:clafeqtrv} for the space component $\order{\bar{E}}{2}_{ij}$, which takes the form
\begin{equation}\label{eqn:tauij2mass}
2\triangle\order{\tau}{2}_{ij} =(c_3 \order{\psi}{2} - c_4 \rho) \delta_{ij}\,,
\end{equation}
where we denoted
\begin{equation}
c_3=\frac{C}{A} m^2_\phi\,, \quad
c_4=\frac{c_\phi C}{A} +\frac{\kappa^2}{A}\,.
\end{equation}
Substituting the solution~\eqref{eqn:psi2solmv} for the second-order scalar field $\order{\psi}{2}$ into the field equation~\eqref{eqn:tauij2mass} we find the solution
\begin{equation}
\order{\tau}{2}_{ij} (r) = \cfrac{M}{8 \pi r} \left[c_3-\frac{c_4 c_\phi}{m^2_\phi} (e^{-m_\phi r} -1) \right] \delta_{ij}\,.
\end{equation}
Again, we find a superposition of a pure Newtonian part and a Yukawa-type term.

\subsection{PPN metric and parameters}\label{ssec:mvmetric}
For the static point source~\eqref{eqn:pointmass} we find that the spherically symmetric post-Newtonian metric is of the general form
\begin{subequations}\label{eqn:ppnmet2}
\begin{align}
g_{00} &=-1+2 \order{\tau}{2}_{00} = -1 + 2G_{\text{eff}}(r)U(r) + \mathcal{O}(4)\,,\label{eqn:g00}\\
g_{0j} &= \mathcal{O}(5)\,,\label{eqn:g0j}\\
g_{ij} &=\delta_{ij}+2 \order{\tau}{2}_{(ij)} = \left[1 + 2G_{\text{eff}}(r)\gamma(r)U(r)\right]\delta_{ij} + \mathcal{O}(4)\,.\label{eqn:gij}
\end{align}
\end{subequations}
Here \(\gamma(r)\) is the post-Newtonian parameter we aim to determine. From our solution for the tetrad we find that it is given by
\begin{equation}\label{eqn:gammamv}
\gamma(r) = \cfrac{2 \omega+3-e^{-m_\phi r}}{2 \omega+3+e^{-m_\phi r}}\,,
\end{equation}
where we introduced $\omega=\cfrac{AB}{C^2}$, while the scalar field mass \(m_{\phi}\) is defined by the relation~\eqref{eqn:scalmass}. We find that the result agrees with the well-known case of a massive scalar field in various scalar-curvature type theories~\cite{Olmo:2005zr,Olmo:2005hc,Perivolaropoulos:2009ak,Hohmann:2013rba,Scharer:2014kya,Hohmann:2015kra}. In particular, it agrees with general relativity in the limit \(\omega \to \infty\) of vanishing kinetic coupling or \(m_{\phi} \to \infty\) of an infinitely heavy scalar field.

\section{Massless case: all PPN parameters}\label{sec:massless}
To proceed further and also solve the perturbative field equations at higher velocity order, we restrict ourselves to theories in which the scalar field is massless, and where its potential satisfies the additional conditions \(V'' = V''' = 0\). It turns out that in this case we can express all perturbations of the tetrad and the spin connection in terms of standard PPN potentials. They are obtained by solving the field equations order by order: the second order scalar field in section~\ref{ssec:mlscal2}, the second order time component of the tetrad in section~\ref{ssec:mltett2}, its second order space components in section~\ref{ssec:mltets2}, its third order components in section~\ref{ssec:mltet3}, and finally the fourth order in section~\ref{ssec:mltett4}. From these solutions we obtain the full metric and post-Newtonian parameters, which we display in section~\ref{ssec:mlmetric}.

\subsection{Scalar field at second order}\label{ssec:mlscal2}
Similarly to the case of a massive scalar field, we start by solving the scalar field equation~\eqref{eqn:debfieldeqns} with~\eqref{eqn:clafeqdeb} at the second velocity order. In the massless case the second-order equation is given by
\begin{equation}
-(2AB + 3C^2)\triangle\order{\psi}{2} = -\kappa^2C\rho\,.
\end{equation}
This is now an ordinary Poisson equation, which has the general solution
\begin{equation}\label{eqn:psi2solml}
\order{\psi}{2} = -\frac{\kappa^2C}{4\pi(2AB + 3C^2)}U
\end{equation}
for an arbitrary source mass density \(\rho\), where the Newtonian potential is defined as the solution of the Poisson equation
\begin{equation}
\triangle U = -4\pi\rho \quad \Leftrightarrow \quad U(t,\vec{x}) = \int d^3x'\frac{\rho(t,\vec{x}')}{|\vec{x} - \vec{x}'|}\,.
\end{equation}
Note that without a mass term, no Yukawa-type dependence arises.

\subsection{Temporal tetrad components at second order}\label{ssec:mltett2}
We then continue with the second-order trace-reversed tetrad equation~\eqref{eqn:trfieldeqns} with~\eqref{eqn:clafeqtrv}. Its time component reads
\begin{equation}
-A\triangle\order{\tau}{2}_{00} - \frac{C}{2}\triangle\order{\psi}{2} = \frac{\kappa^2}{2}\rho\,.
\end{equation}
Using the solution~\eqref{eqn:psi2solml} for the second-order scalar field we thus find the solution
\begin{equation}\label{eqn:tau002ml}
\order{\tau}{2}_{00} = \frac{\kappa^2}{4\pi A}\frac{AB + 2C^2}{2AB + 3C^2}U\,.
\end{equation}
Writing the solution in the form \(\order{\tau}{2}_{00} = GU\), we can read off the gravitational constant
\begin{equation}\label{eqn:gravconst}
G = \frac{\kappa^2}{4\pi A}\frac{AB + 2C^2}{2AB + 3C^2}\,,
\end{equation}
which is now truly a constant and not an effective quantity depending on the distance between source and test mass. Hence, we drop the subscript ``eff''. We will use this expression later to normalize the coupling constant \(\kappa\). Note that also here we find \(G \to \kappa^2/8\pi A\) for \(C \to 0\) as in the massive case~\cite{Jarv:2017npl}.

\subsection{Spatial tetrad components at second order}\label{ssec:mltets2}
In the next step we come to the spatial part of the trace-reversed tetrad equation at the second velocity order, which reads
\begin{equation}\label{eqn:tauij2ml}
A\left(\order{\tau}{2}_{00,ij} - \order{\tau}{2}_{kk,ij} - \triangle\order{\tau}{2}_{(ij)} + \order{\tau}{2}_{(i|k|,j)k} + \order{\tau}{2}_{k(i,j)k}\right) + \frac{C}{2}\left(2\order{\psi}{2}_{,ij} + \triangle\order{\psi}{2}\delta_{ij}\right) = \frac{\kappa^2}{2}\rho\delta_{ij}\,.
\end{equation}
In order to solve this equation, we make use of the diffeomorphism invariance of the theory, which allows us to choose the post-Newtonian coordinate system~\cite{Will:1993ns,Hohmann:2019qgo}, introduce the gauge condition
\begin{equation}\label{eqn:gaugecond1}
0 = K_i = h_{ij,j} - \frac{1}{2}h_{jj,i} + \frac{1}{2}h_{00,i} + \frac{\mathcal{C}}{\mathcal{A}}\psi_{,i}
\end{equation}
on the metric perturbations \(h_{\mu\nu} = g_{\mu\nu} - \eta_{\mu\nu}\), which is a direct adaptation of the gauge condition introduced in~\cite{Nutku:1969} for scalar-curvature gravity. Expanding this gauge condition at the second velocity order and substituting the tetrad perturbations, we find
\begin{equation}
\order{K}{2}_i = \order{\tau}{2}_{ij,j} + \order{\tau}{2}_{ji,j} - \order{\tau}{2}_{jj,i} + \order{\tau}{2}_{00,i} + \frac{C}{A}\order{\psi}{2}_{,i}\,.
\end{equation}
We can implement this gauge condition by adding a suitable multiple to the trace-reversed field equations~\eqref{eqn:trfieldeqns}. Hence, instead of solving the original field equations, we solve the equations
\begin{equation}
\order{\bar{E}}{2}_{ij} - \frac{A}{2}\left(\order{K}{2}_{i,j} + \order{K}{2}_{j,i}\right) = \kappa^2\order{\bar{\Theta}}{2}_{ij}\,,
\end{equation}
which are equivalent to the original equations if the gauge condition~\eqref{eqn:gaugecond1} is satisfied. The second-order tetrad equation then simplifies to
\begin{equation}
-A\triangle\order{\tau}{2}_{(ij)} + \frac{C}{2}\triangle\order{\psi}{2}\delta_{ij} = \frac{\kappa^2}{2}\rho\delta_{ij}\,.
\end{equation}
We thus find the solution
\begin{equation}
\order{\tau}{2}_{(ij)} = \frac{\kappa^2}{4\pi A}\frac{AB + C^2}{2AB + 3C^2}U\delta_{ij}
\end{equation}
for the symmetric part of the spatial tetrad components. Note that the antisymmetric part of the tetrad components is not yet determined by the field equations at this velocity order.

\subsection{Tetrad at third order}\label{ssec:mltet3}
We then continue with the third velocity order. At this stage we need to consider only the symmetric part of the tetrad field equations, which reads
\begin{equation}
A\left(\order{\tau}{2}_{(ij),0j} - \order{\tau}{2}_{jj,0i} - \triangle\order{\tau}{3}_{(0i)} + \order{\tau}{3}_{(0j),ij}\right) + C\order{\psi}{2}_{,0i} = -\kappa^2\rho v_i\,,
\end{equation}
since the remaining equations are satisfied identically. Also in this case we must introduce a gauge condition to fix the post-Newtonian coordinate system~\cite{Will:1993ns,Hohmann:2019qgo}. Here we again follow~\cite{Nutku:1969} and choose the condition
\begin{equation}\label{eqn:gaugecond2}
0 = K_0 = h_{0i,i} - \frac{1}{2}h_{ii,0} + \frac{\mathcal{C}}{\mathcal{A}}\psi_{,0}\,.
\end{equation}
At the third velocity order and with the tetrad perturbations substituted this gauge condition reads
\begin{equation}
\order{K}{3}_0 = \order{\tau}{3}_{0i,i} + \order{\tau}{3}_{i0,i} - \order{\tau}{2}_{ii,0} + \frac{C}{A}\order{\psi}{2}_{,0}\,.
\end{equation}
We then proceed in a similar fashion as for the second velocity order above and subtract a suitable multiple of the gauge condition from the field equations. Hence, the equations we solve are given by
\begin{equation}
\order{\bar{E}}{3}_{(0i)} - \frac{A}{2}\left(\order{C}{3}_{0,i} + \order{C}{2}_{i,0}\right) = \kappa^2\order{\bar{\Theta}}{3}_{(0i)}\,.
\end{equation}
Again we remark that these are equivalent to the original equations, provided that the gauge condition~\eqref{eqn:gaugecond2} is satisfied. We thus obtain the equations
\begin{equation}
-A\triangle\order{\tau}{3}_{(0i)} - \frac{A}{2}\order{\tau}{2}_{00,0i} = -\kappa^2\rho v_i\,.
\end{equation}
Using the previously found solution~\eqref{eqn:tau002ml} we thus find
\begin{equation}
\order{\tau}{3}_{(0i)} = -\frac{\kappa^2}{16\pi A(2AB + 3C^2)}\left[(7AB + 10C^2)V_i + (AB + 2C^2)W_i\right]\,,
\end{equation}
where the post-Newtonian potentials \(V_i\) and \(W_i\) satisfy
\begin{equation}
\triangle V_{i} = -4\pi\rho v_i\,, \quad
\triangle W_{i} = -4\pi\rho v_i + 2 U_{,0i}\,;
\end{equation}
see~\cite[Eq. (4.32)]{Will:1993ns} for their definition. Also in this case we only determine the symmetric part of the tetrad. However, we will see shortly that this will be sufficient for our purpose of determining the post-Newtonian limit.

\subsection{Temporal tetrad components at fourth order}\label{ssec:mltett4}
We finally come to the fourth velocity order, where we must determine the temporal tetrad component. For this purpose we use the temporal component of the fourth-order tetrad field equations, which reads
\begin{equation}
\begin{split}
-A\triangle\order{\tau}{4}_{00} + 2A\order{\tau}{3}_{(0i),0i} - A\order{\tau}{2}_{ii,00} + A\order{\tau}{2}_{00}\triangle\order{\tau}{2}_{00} + 2A\order{\tau}{2}_{ij}\order{\tau}{2}_{00,ij} + 2A\order{\tau}{2}_{00,i}\order{\tau}{2}_{(ij),j}&\\
- A\order{\tau}{2}_{00,i}\order{\tau}{2}_{jj,i} - \frac{C}{2}\triangle\order{\psi}{4} + \frac{3C}{2}\order{\psi}{2}_{,00} + C\order{\tau}{2}_{00}\triangle\order{\psi}{2} + C\order{\tau}{2}_{ij}\order{\psi}{2}_{,ij} - A'\triangle\order{\tau}{2}_{00}\order{\psi}{2}&\\
+ \left(\frac{C}{2} - A'\right)\order{\tau}{2}_{00,i}\order{\psi}{2}_{,i} + C\order{\tau}{2}_{(ij),j}\order{\psi}{2}_{,i} - \frac{C}{2}\order{\tau}{2}_{jj,i}\order{\psi}{2}_{,i} - \frac{C'}{2}\order{\psi}{2}\triangle\order{\psi}{2} - \frac{C'}{2}\order{\psi}{2}_{,i}\order{\psi}{2}_{,i}
&= \kappa^2\rho\left(-\order{\tau}{2}_{00} + v^2 + \frac{\Pi}{2} + \frac{3p}{2\rho}\right)\,.
\end{split}
\end{equation}
In order to eliminate the fourth order scalar field, which appears in form of the term \(\triangle\order{\psi}{4}\), we use the corresponding fourth-order scalar field equation, which is given by
\begin{multline}
(2AB + 3C^2)\left(-\triangle\order{\psi}{4} + \order{\psi}{2}_{,00} + 2\order{\tau}{2}_{ij}\order{\psi}{2}_{,ij} + \order{\tau}{2}_{ij,j}\order{\psi}{2}_{,i}\right) - 2(BA' + AB' + 3CC')\order{\psi}{2}\triangle\order{\psi}{2}\\
+ A(A' + C)\left(4\order{\tau}{2}_{00,i}\order{\tau}{2}_{j[j,i]} + 4\order{\tau}{2}_{i[i,j]}\order{\tau}{2}_{k[j,k]} + \order{\tau}{2}_{(ij),k}\order{\tau}{2}_{(ij),k} + \frac{1}{2}\order{\tau}{2}_{ij,k}\order{\tau}{2}_{kj,i} - \order{\tau}{2}_{ij,k}\order{\tau}{2}_{jk,i} - \frac{1}{2}\order{\tau}{2}_{ij,k}\order{\tau}{2}_{ik,j}\right)\\
+ (BC - AB' - 3CC')\order{\psi}{2}_{,i}\order{\psi}{2}_{,i} + (2AB + C^2 - 2CA')\left(\order{\tau}{2}_{ji,j}\order{\psi}{2}_{,i} - \order{\tau}{2}_{jj,i}\order{\psi}{2}_{,i} + \order{\tau}{2}_{00,i}\order{\psi}{2}_{,i}\right)
= \kappa^2\left[C(3p - \rho\Pi) - C'\order{\psi}{2}\rho\right]\,.
\end{multline}
In the latter equation also appears the antisymmetric part of the second order tetrad, which we obtain by solving the fourth order antisymmetric equation
\begin{equation}
\order{E}{4}_{[ij]} = (A' + C)\left(\order{\tau}{2}_{00,[i}\order{\psi}{2}_{,j]} - \order{\tau}{2}_{kk,[i}\order{\psi}{2}_{,j]} + \order{\tau}{2}_{k[i,|k|}\order{\psi}{2}_{,j]} - \order{\tau}{2}_{k[i,j]}\order{\psi}{2}_{,k}\right) = 0\,,
\end{equation}
which is the lowest order in the perturbative expansion of the antisymmetric equation~\eqref{eqn:clafeqcon}, since its zeroth and second order vanish identically. Also note that the right hand side vanishes, see~\eqref{eqn:fieldeqns}, due to the fact that the energy-momentum tensor derived in~\eqref{eqn:matactvar} is symmetric. Since all remaining terms in this equation arising from scalar fields and symmetric tetrad components are of the form \(U_{,[i}U_{,j]}\) and thus vanish, this is identically solved by setting \(\order{\tau}{2}_{[ij]} = 0\). With the lower order solutions, and following the steps above, one finally obtains for the fourth-order tetrad field equation the form
\begin{equation}
\triangle\order{\tau}{4}_{00} = w_1U_{,00} + w_2U_{,i}U_{,i} + w_3\rho U + w_4\rho\Pi + w_5\rho v^2 + w_6p\,,
\end{equation}
where the constants are given by
\begin{subequations}
\begin{align}
w_1 &= 0\,,\\
w_2 &= -\frac{\kappa^4}{32\pi^2A^2}\frac{4A^3B^3 - 6C^5(A' - 3C) - ABC^3(8A' - 33C) - A^2C[2B^2(A' - 10C) + B'C^2 - 2BCC']}{(2AB + 3C^2)^3}\,,\\
w_3 &= -\frac{\kappa^4}{4\pi A^2}\frac{2A^3B^3 + 6A'C^5 + 2ABC^3(4A' + 3C) + A^2C[B^2(2A' + 7C) + B'C^2 - 2BCC']}{(2AB + 3C^2)^3}\,,\\
w_4 &= -\frac{\kappa^2}{A}\frac{AB + 2C^2}{2AB + 3C^2}\,,\\
w_5 &= -\frac{\kappa^2}{A}\,,\\
w_6 &= -\frac{3\kappa^2}{A}\frac{AB + C^2}{2AB + 3C^2}\,.
\end{align}
\end{subequations}
One recognizes that the terms on the right hand side correspond to the fourth-order post-Newtonian potentials, which satisfy the relations
\begin{equation}
\triangle\Phi_1 = -4\pi\rho v^2\,, \quad
\triangle\Phi_2 = -4\pi\rho U\,, \quad
\triangle\Phi_3 = -4\pi\rho \Pi\,, \quad
\triangle\Phi_4 = -4\pi p\,, \quad
\triangle(\mathscr{A} + \mathscr{B} - \Phi_1) = -2U_{,00}\,,
\end{equation}
where \(\Phi_1, \ldots, \Phi_4, \mathscr{A}, \mathscr{B}\) are defined in~\cite[Eq. (4.35)]{Will:1993ns}. The solution is therefore given by
\begin{equation}
\order{\tau}{4}_{00} = \frac{w_2}{2}U^2 + \left(\frac{w_1}{2} - \frac{w_5}{4\pi}\right)\Phi_1 - \left(w_2 + \frac{w_3}{4\pi}\right)\Phi_2 - \frac{w_4}{4\pi}\Phi_3 - \frac{w_6}{4\pi}\Phi_4 - \frac{w_1}{2}\mathscr{A} - \frac{w_1}{2}\mathscr{B}\,.
\end{equation}
We have thus determined all components of the tetrad which are necessary for calculating the post-Newtonian limit.

\subsection{PPN metric and parameters}\label{ssec:mlmetric}
Using the tetrad components obtained from the calculation detailed above, we can now calculate the post-Newtonian metric components. Using the formula~\eqref{eqn:metricexp} we find the components
\begin{subequations}
\begin{align}
\order{g}{2}_{00} &= 2U\,,\\
\order{g}{2}_{ij} &= 2\frac{AB + C^2}{AB + 2C^2}U\delta_{ij}\,,\\
\order{g}{3}_{0i} &= -\frac{1}{2}\left(\frac{7AB + 10C^2}{AB + 2C^2}V_i + W_i\right)\,,\\
\order{g}{4}_{00} &= \left(3 + \frac{AB}{AB + 2C^2}\right)\Phi_1 + 2\Phi_3 + \left(3 + \frac{3AB}{AB + 2C^2}\right)\Phi_4\nonumber\\
&\phantom{=}+ \frac{8A^3B^3 + 6C^5(A' + 3C) + ABC^3(8A' + 45C) + A^2C[2B^2(A' + 17C) + B'C^2 - 2BCC']}{(AB + 2C^2)^2(2AB + 3C^2)}\Phi_2\nonumber\\
&\phantom{=}- \frac{8A^3B^3 - 6C^5(A' - 7C) - ABC^3(8A' - 73C) - A^2C[2B^2(A' - 21C) + B'C^2 - 2BCC']}{2(AB + 2C^2)^2(2AB + 3C^2)}U^2\,,
\end{align}
\end{subequations}
where we have eliminated \(\kappa\) by using the normalization \(G \equiv 1\) for the gravitational constant~\eqref{eqn:gravconst}. By comparison with the standard PPN metric~\cite[Eq. (4.48)]{Will:1993ns}, which reads
\begin{subequations}
\begin{align}
\order{g}{2}_{00} &= 2U\,,\\
\order{g}{2}_{ij} &= 2\gamma U\delta_{ij}\,,\\
\order{g}{3}_{0i} &= -\frac{1}{2}(3 + 4\gamma + \alpha_1 - \alpha_2 + \zeta_1 - 2 \xi )V_i - \frac{1}{2}(1 + \alpha_2 - \zeta_1 + 2\xi)W_i\,,\\
\order{g}{4}_{00} &= -2\beta U^2 - 2\xi \Phi_W + (2 + 2\gamma + \alpha_3 + \zeta_1  -2\xi)\Phi_1 + 2(1 + 3\gamma - 2\beta + \zeta_2 + \xi)\Phi_2\nonumber\\
&\phantom{=}+ 2(1 + \zeta_3)\Phi_3 + 2(3\gamma + 3\zeta_4 - 2\xi)\Phi_4 - (\zeta_1 - 2\xi) \mathscr{A}\,,
\end{align}
\end{subequations}
we see that this metric is already in the PPN gauge, since \(\mathscr{B}\) does not appear, so that we can immediately read off the PPN parameters. We find that the only non-trivial parameters are given by
\begin{equation}\label{eqn:gammaml}
\gamma = 1 - \frac{C^2}{AB + 2C^2}
\end{equation}
and
\begin{equation}\label{eqn:betaml}
\beta = 1 - \frac{C\{6C^4(C + A') + ABC^2(7C + 8A') + A^2[2B^2(C + A') + B'C^2 - 2BCC']\}}{4(AB + 2C^2)^2(2AB + 3C^2)}\,,
\end{equation}
while all other PPN parameters vanish, \(\xi=\alpha_1=\alpha_2=\alpha_3=\zeta_1=\zeta_2=\zeta_3=\zeta_4=0\). Theories of this type are called fully conservative, as they do not exhibit any preferred-frame or preferred-location effects or violation of the total momentum conservation.

We may also express this result in terms of conformal invariants~\cite{Hohmann:2018ijr}, which are related to the Jordan frame parameter functions we have been using by
\begin{equation}
\mathcal{I}_1 = \frac{1}{\mathcal{A}}\,, \quad
\mathcal{G} = \frac{1}{2}\mathcal{B}\,, \quad
\mathcal{K} = \frac{1}{2}\mathcal{C}\,.
\end{equation}
In terms of these invariant parameter functions the PPN parameters are given by
\begin{equation}
\gamma = 1 - \frac{2K^2I_1}{G + 4K^2I_1}
\end{equation}
and
\begin{equation}
\beta = 1 + \frac{K[(G + 2K^2I_1)(G + 6K^2I_1)I_1' - (2G^2 + 14GK^2I_1 + 24K^4I_1^2 + G'K - 2GK')KI_1^2]}{4I_1(G + 3K^2I_1)(G + 4K^2I_1)^2}
\end{equation}
Note that here \(G\) does \emph{not} denote the gravitational constant, but the constant background value \(G = \mathcal{G}(\Phi)\) of the function \(\mathcal{G}\). We finally remark that the function \(\mathcal{K}\) (or \(\mathcal{C}\) in the Jordan frame, which we used for the calculation), and hence its first Taylor coefficient \(K = \mathcal{K}(\Phi)\), determines the non-minimal kinetic coupling of the scalar field to the teleparallel geometry. We thus see that in the minimal coupling limit \(K \to 0\) both \(\gamma\) and \(\beta\) obtain their general relativity values. Hence, such theories cannot be distinguished from general relativity by their PPN parameters, and more sophisticated methods must be employed to study their phenomenology.

This concludes our derivation of the PPN parameters for the general class of scalar-torsion theories. Particular examples will be discussed in the following section.

\section{Example theories}\label{sec:examples}
In order to further illustrate our results presented in the preceding two sections, we consider a few more specific classes of example theories and calculate their PPN parameters. In section~\ref{ssec:scalcurv} we discuss the teleparallel equivalent of scalar-curvature gravity. Teleparallel dark energy and its generalizations are discussed in section~\ref{ssec:teldarken}. Finally, we discuss theories with a non-minimal coupling to the boundary term in section~\ref{ssec:nonminimal}.

\subsection{Teleparallel equivalent of scalar-curvature gravity}\label{ssec:scalcurv}
For the special case \(\mathcal{C} = -\mathcal{A'}\), it can be shown that the gravitational part of the action reduces to the well-known scalar-tensor gravity action~\cite{Flanagan:2004bz} up to a boundary term, which we neglect here~\cite{Hohmann:2018ijr}:
\begin{equation}\label{eqn:stgactiong}
S_g\left[\theta^a, \omega^a{}_b, \phi\right] = \frac{1}{2\kappa^2}\int_M\left[\mathcal{A}(\phi)\lc{R} + 2\mathcal{B}(\phi)X - 2\kappa^2\mathcal{V}(\phi)\right]\theta\dd^4x\,.
\end{equation}
Substituting \(\mathcal{C}\) by \(-\mathcal{A}'\) we find that our results both in the massive and massless cases indeed reduce to earlier results on the post-Newtonian limit of scalar-curvature gravity~\cite{Nordtvedt:1970uv,Olmo:2005zr,Olmo:2005hc,Perivolaropoulos:2009ak,Jarv:2014hma}.

\subsection{Teleparallel dark energy and its generalizations}\label{ssec:teldarken}
The second class of theories we discuss is conventionally expressed in the Jordan frame \(\alpha \equiv 0\), and its members feature a vanishing kinetic coupling of the scalar field to the torsion, \(\mathcal{C} \equiv 0\), and a general scalar field potential \(\mathcal{V}\). This class features numerous well studied contender theories, which are summarized under the name of (generalized) teleparallel dark energy models, and whose actions are given as follows:
\begin{enumerate}
\item
The classical teleparallel dark energy model~\cite{Geng:2011aj}:
\begin{equation}\label{eqn:teledarken}
S_g = \int_M\left[-\frac{T}{2\kappa^2} + \frac{1}{2}\left(g^{\mu\nu}\phi_{,\mu}\phi_{,\nu} - \xi\phi^2T\right) - \mathcal{V}(\phi)\right]\theta\dd^4x\,,
\end{equation}
with coupling constant \(\xi\) and potential \(\mathcal{V}\). By comparison with the general form~\eqref{eqn:actiong} we find the parameter functions \(\mathcal{A} = 1 + 2\kappa^2\xi\phi^2\) and \(\mathcal{B} = -\kappa^2\).

\item
Interacting dark energy~\cite{Otalora:2013tba}:
\begin{equation}
S_g = \int_M\left[-\frac{T}{2\kappa^2} + \frac{1}{2}\left(g^{\mu\nu}\phi_{,\mu}\phi_{,\nu} - \xi F(\phi)T\right) - \mathcal{V}(\phi)\right]\theta\dd^4x\,,
\end{equation}
where the function \(\mathcal{A}\) is replaced by \(\mathcal{A} = 1 + 2\kappa^2\xi F(\phi)\).

\item
Brans-Dicke type action with a general coupling to torsion~\cite{Izumi:2013dca}:
\begin{equation}
S_g = \int_M\left[-\frac{F(\phi)}{2\kappa^2}T - \omega g^{\mu\nu}\phi_{,\mu}\phi_{,\nu} - \mathcal{V}(\phi)\right]\theta\dd^4x\,,
\end{equation}
where \(\mathcal{A} = F(\phi)\) and \(\mathcal{B} = 2\kappa^2\omega\).

\item
Brans-Dicke type action with a dynamical kinetic term~\cite{Chen:2014qsa}:
\begin{equation}\label{eqn:telebradidynkin}
S_g = \int_M\left[-\frac{\phi}{2\kappa^2}T - \frac{\omega(\phi)}{\phi}g^{\mu\nu}\phi_{,\mu}\phi_{,\nu} - \mathcal{V}(\phi)\right]\theta\dd^4x\,,
\end{equation}
where \(\mathcal{A} = \phi\) and \(\mathcal{B} = 2\kappa^2\omega(\phi)/\phi\).
\end{enumerate}
Due to the fact that all of these models have vanishing kinetic coupling, \(\mathcal{C} = 0\), we find that the PPN parameters we calculated, in the massive case \(\gamma\)~\eqref{eqn:gammamv} and in the massless case \(\gamma\)~\eqref{eqn:gammaml} and \(\beta\)~\eqref{eqn:betaml}, take the values \(\gamma = \beta = 1\). Hence, we conclude that the post-Newtonian limit of these theories agrees with general relativity, so that these theories cannot be distinguished by measurements of the PPN parameters. For the models~\eqref{eqn:teledarken} and~\eqref{eqn:telebradidynkin} our result thus reduces to the PPN parameters found in previous studies~\cite{Li:2013oef,Chen:2014qsa}.

\subsection{Non-minimal coupling to the boundary term}\label{ssec:nonminimal}
The last model we consider employs a non-minimal coupling of the scalar field to the teleparallel boundary term and is defined by the action~\cite{Bahamonde:2015hza}
\begin{equation}
S_g = \int_M\left[-\frac{T}{2\kappa^2} + \frac{1}{2}\left(g^{\mu\nu}\phi_{,\mu}\phi_{,\nu} - \xi\phi^2T - \chi\phi^2B\right) - \mathcal{V}(\phi)\right]\theta\dd^4x
\end{equation}
with constants \(\xi, \chi\) and a general potential \(\mathcal{V}\), and where the boundary term \(B\) is defined via the relation~\eqref{eqn:boundary}. In order to bring the action to the form~\eqref{eqn:actiong} one has to perform integration by parts. After this step one finds the parameter functions
\begin{equation}
\mathcal{A} = 1 + 2\kappa^2\xi\phi^2\,, \quad
\mathcal{B} = -\kappa^2\,, \quad
\mathcal{C} = 4\kappa^2\chi\phi\,.
\end{equation}
Here we restrict ourselves to the massless case \(\mathcal{V} = 0\); see~\cite{Sadjadi:2016kwj} for a discussion of the post-Newtonian limit of the theory with a massive scalar field. Note that the parameter functions explicitly depend on \(\kappa\), so that for the normalization \(G = 1\) of the gravitational constant we must insert them into the expression~\eqref{eqn:gravconst}. This yields the solution
\begin{equation}
\kappa^2 = \frac{16\pi}{1 - 32\pi(\xi - 6\chi^2)\Phi^2 + \sqrt{(1 - 64\pi\chi^2\Phi^2)(1 - 576\pi\chi^2\Phi^2)}}
\end{equation}
as the only solution which yields \(\kappa^2 \to 8\pi\) in the limit \(\Phi \to 0\), as one would expect. Further, observe that \(\mathcal{C} \to 0\) in the limit \(\chi \to 0\). It is thus helpful to expand the PPN parameters in a Taylor series in \(\chi\), since they approach their general relativity values for \(\chi \to 0\). This yields the result
\begin{equation}
\gamma = 1 + 128\pi\chi^2\Phi^2 + \mathcal{O}(\chi^4)\,, \quad
\beta = 1 + 32\pi\xi\chi\Phi^2 + 32\pi\chi^2\Phi^2 + \mathcal{O}(\chi^3)\,.
\end{equation}
Comparison of these results with observations of the PPN parameters thus yields bounds on the appearing constants. However, this would lead beyond the scope of this article.

\section{Conclusion}\label{sec:conclusion}
We have derived the post-Newtonian limit and PPN parameters for a general class of scalar-torsion theories of gravity featuring a non-minimal kinetic coupling between the scalar field and the vector part of the torsion. We found that for the consistency of the post-Newtonian approximation, we must assume a vanishing cosmological background value of the scalar field potential and its first derivative. For the case of a massive scalar field, we calculated the PPN parameter \(\gamma\) under the assumption of a static point mass source. We could drop this assumption in the case of a massless scalar field, for which we calculated the full set of PPN parameters. Our findings show that the class of scalar-torsion theories of gravity is fully conservative in the sense that only the PPN parameters \(\beta\) and \(\gamma\) potentially deviate from their general relativity values, which implies no preferred frame or preferred location effects, as well as the conservation of energy-momentum. We also found that deviating values for \(\beta\) and \(\gamma\) are obtained only for a non-minimal kinetic coupling to vector torsion. Further, we expressed the PPN parameters in terms of quantities which are invariant under conformal transformations.

To illustrate our findings, we applied them to a number of particular models within the class of scalar-torsion theories we considered, and which have been previously considered in the literature mainly as cosmological models. Many of these theories are minimally coupled, and are therefore identical to general relativity at the level of the PPN parameters. We also considered the teleparallel equivalent of scalar-curvature gravity theories, and found that their PPN parameters reproduce the values found for their classical representation through the curvature of the Levi-Civita connection. This confirms the consistency of our approach with previous results.

The work we present here allows for various extensions and generalizations. For example, one may consider more general scalar-torsion theories which include a free function of the torsion and the scalar field~\cite{Hohmann:2018rwf}, and possibly also the scalar field kinetic terms~\cite{Hohmann:2018dqh,Flathmann:2019khc}. This can further be generalized by including more involved kinetic coupling terms to the scalar field, which appear in the teleparallel extension of Horndeski gravity~\cite{Bahamonde:2019shr} or are obtained by applying disformal transformations to the gravitational action~\cite{Hohmann:2019gmt}. Studying these theories would also extend a previous result on the post-Newtonian limit of Horndeski gravity~\cite{Hohmann:2015kra}. Further, one could consider theories with more than one scalar field, which are constructed similarly~\cite{Hohmann:2018ijr}, and calculate their post-Newtonian limit in analogue to the related multi-scalar-curvature theories~\cite{Hohmann:2016yfd}. One might also consider another similarly constructed class of theories, in which the scalar fields couples to the nonmetricity of a likewise flat, but torsion-free connection~\cite{Jarv:2018bgs,Runkla:2018xrv}.

Another possible direction of future research is to calculate higher perturbation orders. There are different approaches which may be pursued. One possibility is to replace the static point mass source we considered by a homogeneous sphere, in order to calculate also the parameter \(\beta\) in the massive case~\cite{Hohmann:2017qje}. Another possibility would be the application of higher order perturbation theory in order to study the emission of gravitational waves during the inspiral phase of a binary black hole merger or similar events~\cite{Blanchet:2013haa}.

\begin{acknowledgments}
EE gratefully acknowledges mobility funding from the European Regional Development Fund through Dora Pluss. MH gratefully acknowledges the full financial support by the Estonian Research Council through the Personal Research Funding project PRG356 and by the European Regional Development Fund through the Center of Excellence TK133 ``The Dark Side of the Universe''.
\end{acknowledgments}

\bibliography{../teleppn}

\begin{thebibliography}{79}%
\makeatletter
\providecommand \@ifxundefined [1]{%
 \@ifx{#1\undefined}
}%
\providecommand \@ifnum [1]{%
 \ifnum #1\expandafter \@firstoftwo
 \else \expandafter \@secondoftwo
 \fi
}%
\providecommand \@ifx [1]{%
 \ifx #1\expandafter \@firstoftwo
 \else \expandafter \@secondoftwo
 \fi
}%
\providecommand \natexlab [1]{#1}%
\providecommand \enquote  [1]{``#1''}%
\providecommand \bibnamefont  [1]{#1}%
\providecommand \bibfnamefont [1]{#1}%
\providecommand \citenamefont [1]{#1}%
\providecommand \href@noop [0]{\@secondoftwo}%
\providecommand \href [0]{\begingroup \@sanitize@url \@href}%
\providecommand \@href[1]{\@@startlink{#1}\@@href}%
\providecommand \@@href[1]{\endgroup#1\@@endlink}%
\providecommand \@sanitize@url [0]{\catcode `\\12\catcode `\$12\catcode
  `\&12\catcode `\#12\catcode `\^12\catcode `\_12\catcode `\%12\relax}%
\providecommand \@@startlink[1]{}%
\providecommand \@@endlink[0]{}%
\providecommand \url  [0]{\begingroup\@sanitize@url \@url }%
\providecommand \@url [1]{\endgroup\@href {#1}{\urlprefix }}%
\providecommand \urlprefix  [0]{URL }%
\providecommand \Eprint [0]{\href }%
\providecommand \doibase [0]{http://dx.doi.org/}%
\providecommand \selectlanguage [0]{\@gobble}%
\providecommand \bibinfo  [0]{\@secondoftwo}%
\providecommand \bibfield  [0]{\@secondoftwo}%
\providecommand \translation [1]{[#1]}%
\providecommand \BibitemOpen [0]{}%
\providecommand \bibitemStop [0]{}%
\providecommand \bibitemNoStop [0]{.\EOS\space}%
\providecommand \EOS [0]{\spacefactor3000\relax}%
\providecommand \BibitemShut  [1]{\csname bibitem#1\endcsname}%
\let\auto@bib@innerbib\@empty
\bibitem [{\citenamefont {Faraoni}(2004)}]{Faraoni:2004pi}%
  \BibitemOpen
  \bibfield  {author} {\bibinfo {author} {\bibfnamefont {Valerio}\ \bibnamefont
  {Faraoni}},\ }\href {\doibase 10.1007/978-1-4020-1989-0} {\emph {\bibinfo
  {title} {{Cosmology in scalar tensor gravity}}}},\ Vol.\ \bibinfo {volume}
  {139}\ (\bibinfo {year} {2004})\BibitemShut {NoStop}%
\bibitem [{\citenamefont {Fujii}\ and\ \citenamefont
  {Maeda}(2007)}]{Fujii:2003pa}%
  \BibitemOpen
  \bibfield  {author} {\bibinfo {author} {\bibfnamefont {Y.}~\bibnamefont
  {Fujii}}\ and\ \bibinfo {author} {\bibfnamefont {K.}~\bibnamefont {Maeda}},\
  }\href {http://www.cambridge.org/uk/catalogue/catalogue.asp?isbn=0521811597}
  {\emph {\bibinfo {title} {{The scalar-tensor theory of gravitation}}}}\
  (\bibinfo  {publisher} {Cambridge University Press},\ \bibinfo {year}
  {2007})\BibitemShut {NoStop}%
\bibitem [{\citenamefont {Flanagan}(2004)}]{Flanagan:2004bz}%
  \BibitemOpen
  \bibfield  {author} {\bibinfo {author} {\bibfnamefont {Eanna~E.}\
  \bibnamefont {Flanagan}},\ }\bibfield  {title} {\enquote {\bibinfo {title}
  {{The Conformal frame freedom in theories of gravitation}},}\ }\href
  {\doibase 10.1088/0264-9381/21/15/N02} {\bibfield  {journal} {\bibinfo
  {journal} {Class. Quant. Grav.}\ }\textbf {\bibinfo {volume} {21}},\ \bibinfo
  {pages} {3817} (\bibinfo {year} {2004})},\ \Eprint
  {http://arxiv.org/abs/gr-qc/0403063} {arXiv:gr-qc/0403063 [gr-qc]}
  \BibitemShut {NoStop}%
\bibitem [{\citenamefont {Capozziello}\ \emph {et~al.}(2006)\citenamefont
  {Capozziello}, \citenamefont {Nojiri}, \citenamefont {Odintsov},\ and\
  \citenamefont {Troisi}}]{Capozziello:2006dj}%
  \BibitemOpen
  \bibfield  {author} {\bibinfo {author} {\bibfnamefont {Salvatore}\
  \bibnamefont {Capozziello}}, \bibinfo {author} {\bibfnamefont
  {S.}~\bibnamefont {Nojiri}}, \bibinfo {author} {\bibfnamefont {S.~D.}\
  \bibnamefont {Odintsov}}, \ and\ \bibinfo {author} {\bibfnamefont
  {A.}~\bibnamefont {Troisi}},\ }\bibfield  {title} {\enquote {\bibinfo {title}
  {{Cosmological viability of f(R)-gravity as an ideal fluid and its
  compatibility with a matter dominated phase}},}\ }\href {\doibase
  10.1016/j.physletb.2006.06.034} {\bibfield  {journal} {\bibinfo  {journal}
  {Phys. Lett.}\ }\textbf {\bibinfo {volume} {B639}},\ \bibinfo {pages}
  {135--143} (\bibinfo {year} {2006})},\ \Eprint
  {http://arxiv.org/abs/astro-ph/0604431} {arXiv:astro-ph/0604431 [astro-ph]}
  \BibitemShut {NoStop}%
\bibitem [{\citenamefont {Catena}\ \emph {et~al.}(2007)\citenamefont {Catena},
  \citenamefont {Pietroni},\ and\ \citenamefont {Scarabello}}]{Catena:2006bd}%
  \BibitemOpen
  \bibfield  {author} {\bibinfo {author} {\bibfnamefont {Riccardo}\
  \bibnamefont {Catena}}, \bibinfo {author} {\bibfnamefont {Massimo}\
  \bibnamefont {Pietroni}}, \ and\ \bibinfo {author} {\bibfnamefont {Luca}\
  \bibnamefont {Scarabello}},\ }\bibfield  {title} {\enquote {\bibinfo {title}
  {{Einstein and Jordan reconciled: a frame-invariant approach to scalar-tensor
  cosmology}},}\ }\href {\doibase 10.1103/PhysRevD.76.084039} {\bibfield
  {journal} {\bibinfo  {journal} {Phys. Rev.}\ }\textbf {\bibinfo {volume}
  {D76}},\ \bibinfo {pages} {084039} (\bibinfo {year} {2007})},\ \Eprint
  {http://arxiv.org/abs/astro-ph/0604492} {arXiv:astro-ph/0604492 [astro-ph]}
  \BibitemShut {NoStop}%
\bibitem [{\citenamefont {Faraoni}\ and\ \citenamefont
  {Nadeau}(2007)}]{Faraoni:2006fx}%
  \BibitemOpen
  \bibfield  {author} {\bibinfo {author} {\bibfnamefont {Valerio}\ \bibnamefont
  {Faraoni}}\ and\ \bibinfo {author} {\bibfnamefont {Shahn}\ \bibnamefont
  {Nadeau}},\ }\bibfield  {title} {\enquote {\bibinfo {title} {{The
  (pseudo)issue of the conformal frame revisited}},}\ }\href {\doibase
  10.1103/PhysRevD.75.023501} {\bibfield  {journal} {\bibinfo  {journal} {Phys.
  Rev.}\ }\textbf {\bibinfo {volume} {D75}},\ \bibinfo {pages} {023501}
  (\bibinfo {year} {2007})},\ \Eprint {http://arxiv.org/abs/gr-qc/0612075}
  {arXiv:gr-qc/0612075 [gr-qc]} \BibitemShut {NoStop}%
\bibitem [{\citenamefont {Deruelle}\ and\ \citenamefont
  {Sasaki}(2011)}]{Deruelle:2010ht}%
  \BibitemOpen
  \bibfield  {author} {\bibinfo {author} {\bibfnamefont {Nathalie}\
  \bibnamefont {Deruelle}}\ and\ \bibinfo {author} {\bibfnamefont {Misao}\
  \bibnamefont {Sasaki}},\ }\bibfield  {title} {\enquote {\bibinfo {title}
  {{Conformal equivalence in classical gravity: the example of 'Veiled' General
  Relativity}},}\ }\bibfield  {booktitle} {\emph {\bibinfo {booktitle}
  {{Proceedings, Cosmology, the Quantum Vacuum, and Zeta Functions: Bellaterra,
  Barcelona, Spain, March 8-10, 2010}}},\ }\href {\doibase
  10.1007/978-3-642-19760-4_23} {\bibfield  {journal} {\bibinfo  {journal}
  {Springer Proc. Phys.}\ }\textbf {\bibinfo {volume} {137}},\ \bibinfo {pages}
  {247--260} (\bibinfo {year} {2011})},\ \Eprint
  {http://arxiv.org/abs/1007.3563} {arXiv:1007.3563 [gr-qc]} \BibitemShut
  {NoStop}%
\bibitem [{\citenamefont {Chiba}\ and\ \citenamefont
  {Yamaguchi}(2013)}]{Chiba:2013mha}%
  \BibitemOpen
  \bibfield  {author} {\bibinfo {author} {\bibfnamefont {Takeshi}\ \bibnamefont
  {Chiba}}\ and\ \bibinfo {author} {\bibfnamefont {Masahide}\ \bibnamefont
  {Yamaguchi}},\ }\bibfield  {title} {\enquote {\bibinfo {title}
  {{Conformal-Frame (In)dependence of Cosmological Observations in
  Scalar-Tensor Theory}},}\ }\href {\doibase 10.1088/1475-7516/2013/10/040}
  {\bibfield  {journal} {\bibinfo  {journal} {JCAP}\ }\textbf {\bibinfo
  {volume} {1310}},\ \bibinfo {pages} {040} (\bibinfo {year} {2013})},\ \Eprint
  {http://arxiv.org/abs/1308.1142} {arXiv:1308.1142 [gr-qc]} \BibitemShut
  {NoStop}%
\bibitem [{\citenamefont {Postma}\ and\ \citenamefont
  {Volponi}(2014)}]{Postma:2014vaa}%
  \BibitemOpen
  \bibfield  {author} {\bibinfo {author} {\bibfnamefont {Marieke}\ \bibnamefont
  {Postma}}\ and\ \bibinfo {author} {\bibfnamefont {Marco}\ \bibnamefont
  {Volponi}},\ }\bibfield  {title} {\enquote {\bibinfo {title} {{Equivalence of
  the Einstein and Jordan frames}},}\ }\href {\doibase
  10.1103/PhysRevD.90.103516} {\bibfield  {journal} {\bibinfo  {journal} {Phys.
  Rev.}\ }\textbf {\bibinfo {volume} {D90}},\ \bibinfo {pages} {103516}
  (\bibinfo {year} {2014})},\ \Eprint {http://arxiv.org/abs/1407.6874}
  {arXiv:1407.6874 [astro-ph.CO]} \BibitemShut {NoStop}%
\bibitem [{\citenamefont {Faraoni}\ \emph {et~al.}(1999)\citenamefont
  {Faraoni}, \citenamefont {Gunzig},\ and\ \citenamefont
  {Nardone}}]{Faraoni:1998qx}%
  \BibitemOpen
  \bibfield  {author} {\bibinfo {author} {\bibfnamefont {Valerio}\ \bibnamefont
  {Faraoni}}, \bibinfo {author} {\bibfnamefont {Edgard}\ \bibnamefont
  {Gunzig}}, \ and\ \bibinfo {author} {\bibfnamefont {Pasquale}\ \bibnamefont
  {Nardone}},\ }\bibfield  {title} {\enquote {\bibinfo {title} {{Conformal
  transformations in classical gravitational theories and in cosmology}},}\
  }\href@noop {} {\bibfield  {journal} {\bibinfo  {journal} {Fund. Cosmic
  Phys.}\ }\textbf {\bibinfo {volume} {20}},\ \bibinfo {pages} {121} (\bibinfo
  {year} {1999})},\ \Eprint {http://arxiv.org/abs/gr-qc/9811047}
  {arXiv:gr-qc/9811047 [gr-qc]} \BibitemShut {NoStop}%
\bibitem [{\citenamefont {Capozziello}\ \emph {et~al.}(2010)\citenamefont
  {Capozziello}, \citenamefont {Martin-Moruno},\ and\ \citenamefont
  {Rubano}}]{Capozziello:2010sc}%
  \BibitemOpen
  \bibfield  {author} {\bibinfo {author} {\bibfnamefont {S.}~\bibnamefont
  {Capozziello}}, \bibinfo {author} {\bibfnamefont {P.}~\bibnamefont
  {Martin-Moruno}}, \ and\ \bibinfo {author} {\bibfnamefont {C.}~\bibnamefont
  {Rubano}},\ }\bibfield  {title} {\enquote {\bibinfo {title} {{Physical
  non-equivalence of the Jordan and Einstein frames}},}\ }\href {\doibase
  10.1016/j.physletb.2010.04.058} {\bibfield  {journal} {\bibinfo  {journal}
  {Phys. Lett.}\ }\textbf {\bibinfo {volume} {B689}},\ \bibinfo {pages}
  {117--121} (\bibinfo {year} {2010})},\ \Eprint
  {http://arxiv.org/abs/1003.5394} {arXiv:1003.5394 [gr-qc]} \BibitemShut
  {NoStop}%
\bibitem [{\citenamefont {Rondeau}\ and\ \citenamefont
  {Li}(2017)}]{Rondeau:2017xck}%
  \BibitemOpen
  \bibfield  {author} {\bibinfo {author} {\bibfnamefont {François}\
  \bibnamefont {Rondeau}}\ and\ \bibinfo {author} {\bibfnamefont {Baojiu}\
  \bibnamefont {Li}},\ }\bibfield  {title} {\enquote {\bibinfo {title}
  {{Equivalence of cosmological observables in conformally related scalar
  tensor theories}},}\ }\href {\doibase 10.1103/PhysRevD.96.124009} {\bibfield
  {journal} {\bibinfo  {journal} {Phys. Rev.}\ }\textbf {\bibinfo {volume}
  {D96}},\ \bibinfo {pages} {124009} (\bibinfo {year} {2017})},\ \Eprint
  {http://arxiv.org/abs/1709.07087} {arXiv:1709.07087 [gr-qc]} \BibitemShut
  {NoStop}%
\bibitem [{\citenamefont {Bahamonde}\ \emph {et~al.}(2016)\citenamefont
  {Bahamonde}, \citenamefont {Odintsov}, \citenamefont {Oikonomou},\ and\
  \citenamefont {Wright}}]{Bahamonde:2016wmz}%
  \BibitemOpen
  \bibfield  {author} {\bibinfo {author} {\bibfnamefont {Sebastian}\
  \bibnamefont {Bahamonde}}, \bibinfo {author} {\bibfnamefont {S.~D.}\
  \bibnamefont {Odintsov}}, \bibinfo {author} {\bibfnamefont {V.~K.}\
  \bibnamefont {Oikonomou}}, \ and\ \bibinfo {author} {\bibfnamefont {Matthew}\
  \bibnamefont {Wright}},\ }\bibfield  {title} {\enquote {\bibinfo {title}
  {{Correspondence of $F(R)$ Gravity Singularities in Jordan and Einstein
  Frames}},}\ }\href {\doibase 10.1016/j.aop.2016.06.020} {\bibfield  {journal}
  {\bibinfo  {journal} {Annals Phys.}\ }\textbf {\bibinfo {volume} {373}},\
  \bibinfo {pages} {96--114} (\bibinfo {year} {2016})},\ \Eprint
  {http://arxiv.org/abs/1603.05113} {arXiv:1603.05113 [gr-qc]} \BibitemShut
  {NoStop}%
\bibitem [{\citenamefont {Bahamonde}\ \emph {et~al.}(2017)\citenamefont
  {Bahamonde}, \citenamefont {Odintsov}, \citenamefont {Oikonomou},\ and\
  \citenamefont {Tretyakov}}]{Bahamonde:2017kbs}%
  \BibitemOpen
  \bibfield  {author} {\bibinfo {author} {\bibfnamefont {Sebastian}\
  \bibnamefont {Bahamonde}}, \bibinfo {author} {\bibfnamefont {Sergei~D.}\
  \bibnamefont {Odintsov}}, \bibinfo {author} {\bibfnamefont {V.~K.}\
  \bibnamefont {Oikonomou}}, \ and\ \bibinfo {author} {\bibfnamefont {Petr~V.}\
  \bibnamefont {Tretyakov}},\ }\bibfield  {title} {\enquote {\bibinfo {title}
  {{Deceleration versus acceleration universe in different frames of $F(R)$
  gravity}},}\ }\href {\doibase 10.1016/j.physletb.2017.01.012} {\bibfield
  {journal} {\bibinfo  {journal} {Phys. Lett.}\ }\textbf {\bibinfo {volume}
  {B766}},\ \bibinfo {pages} {225--230} (\bibinfo {year} {2017})},\ \Eprint
  {http://arxiv.org/abs/1701.02381} {arXiv:1701.02381 [gr-qc]} \BibitemShut
  {NoStop}%
\bibitem [{\citenamefont {Brooker}\ \emph {et~al.}(2016)\citenamefont
  {Brooker}, \citenamefont {Odintsov},\ and\ \citenamefont
  {Woodard}}]{Brooker:2016oqa}%
  \BibitemOpen
  \bibfield  {author} {\bibinfo {author} {\bibfnamefont {D.~J.}\ \bibnamefont
  {Brooker}}, \bibinfo {author} {\bibfnamefont {S.~D.}\ \bibnamefont
  {Odintsov}}, \ and\ \bibinfo {author} {\bibfnamefont {R.~P.}\ \bibnamefont
  {Woodard}},\ }\bibfield  {title} {\enquote {\bibinfo {title} {{Precision
  predictions for the primordial power spectra from $f(R)$ models of
  inflation}},}\ }\href {\doibase 10.1016/j.nuclphysb.2016.08.010} {\bibfield
  {journal} {\bibinfo  {journal} {Nucl. Phys.}\ }\textbf {\bibinfo {volume}
  {B911}},\ \bibinfo {pages} {318--337} (\bibinfo {year} {2016})},\ \Eprint
  {http://arxiv.org/abs/1606.05879} {arXiv:1606.05879 [gr-qc]} \BibitemShut
  {NoStop}%
\bibitem [{\citenamefont {Järv}\ \emph {et~al.}(2015)\citenamefont {Järv},
  \citenamefont {Kuusk}, \citenamefont {Saal},\ and\ \citenamefont
  {Vilson}}]{Jarv:2014hma}%
  \BibitemOpen
  \bibfield  {author} {\bibinfo {author} {\bibfnamefont {Laur}\ \bibnamefont
  {Järv}}, \bibinfo {author} {\bibfnamefont {Piret}\ \bibnamefont {Kuusk}},
  \bibinfo {author} {\bibfnamefont {Margus}\ \bibnamefont {Saal}}, \ and\
  \bibinfo {author} {\bibfnamefont {Ott}\ \bibnamefont {Vilson}},\ }\bibfield
  {title} {\enquote {\bibinfo {title} {{Invariant quantities in the
  scalar-tensor theories of gravitation}},}\ }\href {\doibase
  10.1103/PhysRevD.91.024041} {\bibfield  {journal} {\bibinfo  {journal} {Phys.
  Rev.}\ }\textbf {\bibinfo {volume} {D91}},\ \bibinfo {pages} {024041}
  (\bibinfo {year} {2015})},\ \Eprint {http://arxiv.org/abs/1411.1947}
  {arXiv:1411.1947 [gr-qc]} \BibitemShut {NoStop}%
\bibitem [{\citenamefont {Kuusk}\ \emph {et~al.}(2016)\citenamefont {Kuusk},
  \citenamefont {Jarv},\ and\ \citenamefont {Vilson}}]{Kuusk:2015dda}%
  \BibitemOpen
  \bibfield  {author} {\bibinfo {author} {\bibfnamefont {Piret}\ \bibnamefont
  {Kuusk}}, \bibinfo {author} {\bibfnamefont {Laur}\ \bibnamefont {Jarv}}, \
  and\ \bibinfo {author} {\bibfnamefont {Ott}\ \bibnamefont {Vilson}},\
  }\bibfield  {title} {\enquote {\bibinfo {title} {{Invariant quantities in the
  multiscalar-tensor theories of gravitation}},}\ }\href {\doibase
  10.1142/S0217751X16410037} {\bibfield  {journal} {\bibinfo  {journal} {Int.
  J. Mod. Phys.}\ }\textbf {\bibinfo {volume} {A31}},\ \bibinfo {pages}
  {1641003} (\bibinfo {year} {2016})},\ \Eprint
  {http://arxiv.org/abs/1509.02903} {arXiv:1509.02903 [gr-qc]} \BibitemShut
  {NoStop}%
\bibitem [{\citenamefont {Jiménez}\ \emph
  {et~al.}(2019{\natexlab{a}})\citenamefont {Jiménez}, \citenamefont
  {Heisenberg},\ and\ \citenamefont {Koivisto}}]{BeltranJimenez:2019tjy}%
  \BibitemOpen
  \bibfield  {author} {\bibinfo {author} {\bibfnamefont {Jose~Beltrán}\
  \bibnamefont {Jiménez}}, \bibinfo {author} {\bibfnamefont {Lavinia}\
  \bibnamefont {Heisenberg}}, \ and\ \bibinfo {author} {\bibfnamefont
  {Tomi~S.}\ \bibnamefont {Koivisto}},\ }\bibfield  {title} {\enquote {\bibinfo
  {title} {{The Geometrical Trinity of Gravity}},}\ }\href {\doibase
  10.3390/universe5070173} {\bibfield  {journal} {\bibinfo  {journal}
  {Universe}\ }\textbf {\bibinfo {volume} {5}},\ \bibinfo {pages} {173}
  (\bibinfo {year} {2019}{\natexlab{a}})},\ \Eprint
  {http://arxiv.org/abs/1903.06830} {arXiv:1903.06830 [hep-th]} \BibitemShut
  {NoStop}%
\bibitem [{\citenamefont {Jiménez}\ \emph
  {et~al.}(2019{\natexlab{b}})\citenamefont {Jiménez}, \citenamefont
  {Heisenberg}, \citenamefont {Iosifidis}, \citenamefont {Jiménez-Cano},\ and\
  \citenamefont {Koivisto}}]{Jimenez:2019ghw}%
  \BibitemOpen
  \bibfield  {author} {\bibinfo {author} {\bibfnamefont {Jose~Beltrán}\
  \bibnamefont {Jiménez}}, \bibinfo {author} {\bibfnamefont {Lavinia}\
  \bibnamefont {Heisenberg}}, \bibinfo {author} {\bibfnamefont {Damianos}\
  \bibnamefont {Iosifidis}}, \bibinfo {author} {\bibfnamefont {Alejandro}\
  \bibnamefont {Jiménez-Cano}}, \ and\ \bibinfo {author} {\bibfnamefont
  {Tomi~S.}\ \bibnamefont {Koivisto}},\ }\bibfield  {title} {\enquote {\bibinfo
  {title} {{General Teleparallel Quadratic Gravity}},}\ }\href@noop {} {\
  (\bibinfo {year} {2019}{\natexlab{b}})},\ \Eprint
  {http://arxiv.org/abs/1909.09045} {arXiv:1909.09045 [gr-qc]} \BibitemShut
  {NoStop}%
\bibitem [{\citenamefont {Einstein}(1928)}]{Einstein:1928}%
  \BibitemOpen
  \bibfield  {author} {\bibinfo {author} {\bibfnamefont {Albert}\ \bibnamefont
  {Einstein}},\ }\bibfield  {title} {\enquote {\bibinfo {title}
  {{Riemann-Geometrie mit Aufrechterhaltung des Begriffes des
  Fernparallelismus}},}\ }\href
  {http://echo.mpiwg-berlin.mpg.de/MPIWG:YP5DFQU1} {\bibfield  {journal}
  {\bibinfo  {journal} {Sitzber. Preuss. Akad. Wiss.}\ }\textbf {\bibinfo
  {volume} {17}},\ \bibinfo {pages} {217--221} (\bibinfo {year}
  {1928})}\BibitemShut {NoStop}%
\bibitem [{\citenamefont {M{\o}ller}(1961)}]{Moller:1961}%
  \BibitemOpen
  \bibfield  {author} {\bibinfo {author} {\bibfnamefont {Christian}\
  \bibnamefont {M{\o}ller}},\ }\bibfield  {title} {\enquote {\bibinfo {title}
  {{Conservation Laws and Absolute Parallelism in General Relativity}},}\
  }\href@noop {} {\bibfield  {journal} {\bibinfo  {journal} {K. Dan. Vidensk.
  Selsk. Mat. Fys. Skr.}\ }\textbf {\bibinfo {volume} {1}},\ \bibinfo {pages}
  {1--50} (\bibinfo {year} {1961})}\BibitemShut {NoStop}%
\bibitem [{\citenamefont {Aldrovandi}\ and\ \citenamefont
  {Pereira}(2013)}]{Aldrovandi:2013wha}%
  \BibitemOpen
  \bibfield  {author} {\bibinfo {author} {\bibfnamefont {Ruben}\ \bibnamefont
  {Aldrovandi}}\ and\ \bibinfo {author} {\bibfnamefont {José~Geraldo}\
  \bibnamefont {Pereira}},\ }\href {\doibase 10.1007/978-94-007-5143-9} {\emph
  {\bibinfo {title} {{Teleparallel Gravity}}}},\ Vol.\ \bibinfo {volume} {173}\
  (\bibinfo  {publisher} {Springer},\ \bibinfo {address} {Dordrecht},\ \bibinfo
  {year} {2013})\BibitemShut {NoStop}%
\bibitem [{\citenamefont {Maluf}(2013)}]{Maluf:2013gaa}%
  \BibitemOpen
  \bibfield  {author} {\bibinfo {author} {\bibfnamefont {J.~W.}\ \bibnamefont
  {Maluf}},\ }\bibfield  {title} {\enquote {\bibinfo {title} {{The teleparallel
  equivalent of general relativity}},}\ }\href {\doibase
  10.1002/andp.201200272} {\bibfield  {journal} {\bibinfo  {journal} {Annalen
  Phys.}\ }\textbf {\bibinfo {volume} {525}},\ \bibinfo {pages} {339--357}
  (\bibinfo {year} {2013})},\ \Eprint {http://arxiv.org/abs/1303.3897}
  {arXiv:1303.3897 [gr-qc]} \BibitemShut {NoStop}%
\bibitem [{\citenamefont {Golovnev}(2018)}]{Golovnev:2018red}%
  \BibitemOpen
  \bibfield  {author} {\bibinfo {author} {\bibfnamefont {Alexey}\ \bibnamefont
  {Golovnev}},\ }\bibfield  {title} {\enquote {\bibinfo {title} {{Introduction
  to teleparallel gravities}},}\ }in\ \href
  {http://inspirehep.net/record/1649207/files/arXiv:1801.06929.pdf} {\emph
  {\bibinfo {booktitle} {{9th Mathematical Physics Meeting: Summer School and
  Conference on Modern Mathematical Physics Belgrade, Serbia, September 18-23,
  2017}}}}\ (\bibinfo {year} {2018})\ \Eprint {http://arxiv.org/abs/1801.06929}
  {arXiv:1801.06929 [gr-qc]} \BibitemShut {NoStop}%
\bibitem [{\citenamefont {Li}\ \emph {et~al.}(2011{\natexlab{a}})\citenamefont
  {Li}, \citenamefont {Sotiriou},\ and\ \citenamefont {Barrow}}]{Li:2010cg}%
  \BibitemOpen
  \bibfield  {author} {\bibinfo {author} {\bibfnamefont {Baojiu}\ \bibnamefont
  {Li}}, \bibinfo {author} {\bibfnamefont {Thomas~P.}\ \bibnamefont
  {Sotiriou}}, \ and\ \bibinfo {author} {\bibfnamefont {John~D.}\ \bibnamefont
  {Barrow}},\ }\bibfield  {title} {\enquote {\bibinfo {title} {{$f(T)$ gravity
  and local Lorentz invariance}},}\ }\href {\doibase
  10.1103/PhysRevD.83.064035} {\bibfield  {journal} {\bibinfo  {journal} {Phys.
  Rev.}\ }\textbf {\bibinfo {volume} {D83}},\ \bibinfo {pages} {064035}
  (\bibinfo {year} {2011}{\natexlab{a}})},\ \Eprint
  {http://arxiv.org/abs/1010.1041} {arXiv:1010.1041 [gr-qc]} \BibitemShut
  {NoStop}%
\bibitem [{\citenamefont {Sotiriou}\ \emph {et~al.}(2011)\citenamefont
  {Sotiriou}, \citenamefont {Li},\ and\ \citenamefont
  {Barrow}}]{Sotiriou:2010mv}%
  \BibitemOpen
  \bibfield  {author} {\bibinfo {author} {\bibfnamefont {Thomas~P.}\
  \bibnamefont {Sotiriou}}, \bibinfo {author} {\bibfnamefont {Baojiu}\
  \bibnamefont {Li}}, \ and\ \bibinfo {author} {\bibfnamefont {John~D.}\
  \bibnamefont {Barrow}},\ }\bibfield  {title} {\enquote {\bibinfo {title}
  {{Generalizations of teleparallel gravity and local Lorentz symmetry}},}\
  }\href {\doibase 10.1103/PhysRevD.83.104030} {\bibfield  {journal} {\bibinfo
  {journal} {Phys. Rev.}\ }\textbf {\bibinfo {volume} {D83}},\ \bibinfo {pages}
  {104030} (\bibinfo {year} {2011})},\ \Eprint {http://arxiv.org/abs/1012.4039}
  {arXiv:1012.4039 [gr-qc]} \BibitemShut {NoStop}%
\bibitem [{\citenamefont {Li}\ \emph {et~al.}(2011{\natexlab{b}})\citenamefont
  {Li}, \citenamefont {Miao},\ and\ \citenamefont {Miao}}]{Li:2011rn}%
  \BibitemOpen
  \bibfield  {author} {\bibinfo {author} {\bibfnamefont {Miao}\ \bibnamefont
  {Li}}, \bibinfo {author} {\bibfnamefont {Rong-Xin}\ \bibnamefont {Miao}}, \
  and\ \bibinfo {author} {\bibfnamefont {Yan-Gang}\ \bibnamefont {Miao}},\
  }\bibfield  {title} {\enquote {\bibinfo {title} {{Degrees of freedom of
  $f(T)$ gravity}},}\ }\href {\doibase 10.1007/JHEP07(2011)108} {\bibfield
  {journal} {\bibinfo  {journal} {JHEP}\ }\textbf {\bibinfo {volume} {07}},\
  \bibinfo {pages} {108} (\bibinfo {year} {2011}{\natexlab{b}})},\ \Eprint
  {http://arxiv.org/abs/1105.5934} {arXiv:1105.5934 [hep-th]} \BibitemShut
  {NoStop}%
\bibitem [{\citenamefont {Ong}\ \emph {et~al.}(2013)\citenamefont {Ong},
  \citenamefont {Izumi}, \citenamefont {Nester},\ and\ \citenamefont
  {Chen}}]{Ong:2013qja}%
  \BibitemOpen
  \bibfield  {author} {\bibinfo {author} {\bibfnamefont {Yen~Chin}\
  \bibnamefont {Ong}}, \bibinfo {author} {\bibfnamefont {Keisuke}\ \bibnamefont
  {Izumi}}, \bibinfo {author} {\bibfnamefont {James~M.}\ \bibnamefont
  {Nester}}, \ and\ \bibinfo {author} {\bibfnamefont {Pisin}\ \bibnamefont
  {Chen}},\ }\bibfield  {title} {\enquote {\bibinfo {title} {{Problems with
  Propagation and Time Evolution in f(T) Gravity}},}\ }\href {\doibase
  10.1103/PhysRevD.88.024019} {\bibfield  {journal} {\bibinfo  {journal} {Phys.
  Rev.}\ }\textbf {\bibinfo {volume} {D88}},\ \bibinfo {pages} {024019}
  (\bibinfo {year} {2013})},\ \Eprint {http://arxiv.org/abs/1303.0993}
  {arXiv:1303.0993 [gr-qc]} \BibitemShut {NoStop}%
\bibitem [{\citenamefont {Izumi}\ \emph {et~al.}(2014)\citenamefont {Izumi},
  \citenamefont {Gu},\ and\ \citenamefont {Ong}}]{Izumi:2013dca}%
  \BibitemOpen
  \bibfield  {author} {\bibinfo {author} {\bibfnamefont {Keisuke}\ \bibnamefont
  {Izumi}}, \bibinfo {author} {\bibfnamefont {Je-An}\ \bibnamefont {Gu}}, \
  and\ \bibinfo {author} {\bibfnamefont {Yen~Chin}\ \bibnamefont {Ong}},\
  }\bibfield  {title} {\enquote {\bibinfo {title} {{Acausality and Nonunique
  Evolution in Generalized Teleparallel Gravity}},}\ }\href {\doibase
  10.1103/PhysRevD.89.084025} {\bibfield  {journal} {\bibinfo  {journal} {Phys.
  Rev.}\ }\textbf {\bibinfo {volume} {D89}},\ \bibinfo {pages} {084025}
  (\bibinfo {year} {2014})},\ \Eprint {http://arxiv.org/abs/1309.6461}
  {arXiv:1309.6461 [gr-qc]} \BibitemShut {NoStop}%
\bibitem [{\citenamefont {Chen}\ \emph
  {et~al.}(2015{\natexlab{a}})\citenamefont {Chen}, \citenamefont {Izumi},
  \citenamefont {Nester},\ and\ \citenamefont {Ong}}]{Chen:2014qtl}%
  \BibitemOpen
  \bibfield  {author} {\bibinfo {author} {\bibfnamefont {Pisin}\ \bibnamefont
  {Chen}}, \bibinfo {author} {\bibfnamefont {Keisuke}\ \bibnamefont {Izumi}},
  \bibinfo {author} {\bibfnamefont {James~M.}\ \bibnamefont {Nester}}, \ and\
  \bibinfo {author} {\bibfnamefont {Yen~Chin}\ \bibnamefont {Ong}},\ }\bibfield
   {title} {\enquote {\bibinfo {title} {{Remnant Symmetry, Propagation and
  Evolution in $f$(T) Gravity}},}\ }\href {\doibase 10.1103/PhysRevD.91.064003}
  {\bibfield  {journal} {\bibinfo  {journal} {Phys. Rev.}\ }\textbf {\bibinfo
  {volume} {D91}},\ \bibinfo {pages} {064003} (\bibinfo {year}
  {2015}{\natexlab{a}})},\ \Eprint {http://arxiv.org/abs/1412.8383}
  {arXiv:1412.8383 [gr-qc]} \BibitemShut {NoStop}%
\bibitem [{\citenamefont {Beltrán~Jiménez}\ \emph {et~al.}(2018)\citenamefont
  {Beltrán~Jiménez}, \citenamefont {Heisenberg},\ and\ \citenamefont
  {Koivisto}}]{BeltranJimenez:2018vdo}%
  \BibitemOpen
  \bibfield  {author} {\bibinfo {author} {\bibfnamefont {Jose}\ \bibnamefont
  {Beltrán~Jiménez}}, \bibinfo {author} {\bibfnamefont {Lavinia}\
  \bibnamefont {Heisenberg}}, \ and\ \bibinfo {author} {\bibfnamefont
  {Tomi~S.}\ \bibnamefont {Koivisto}},\ }\bibfield  {title} {\enquote {\bibinfo
  {title} {{Teleparallel Palatini theories}},}\ }\href {\doibase
  10.1088/1475-7516/2018/08/039} {\bibfield  {journal} {\bibinfo  {journal}
  {JCAP}\ }\textbf {\bibinfo {volume} {1808}},\ \bibinfo {pages} {039}
  (\bibinfo {year} {2018})},\ \Eprint {http://arxiv.org/abs/1803.10185}
  {arXiv:1803.10185 [gr-qc]} \BibitemShut {NoStop}%
\bibitem [{\citenamefont {Krššák}\ and\ \citenamefont
  {Saridakis}(2016)}]{Krssak:2015oua}%
  \BibitemOpen
  \bibfield  {author} {\bibinfo {author} {\bibfnamefont {Martin}\ \bibnamefont
  {Krššák}}\ and\ \bibinfo {author} {\bibfnamefont {Emmanuel~N.}\
  \bibnamefont {Saridakis}},\ }\bibfield  {title} {\enquote {\bibinfo {title}
  {{The covariant formulation of f(T) gravity}},}\ }\href {\doibase
  10.1088/0264-9381/33/11/115009} {\bibfield  {journal} {\bibinfo  {journal}
  {Class. Quant. Grav.}\ }\textbf {\bibinfo {volume} {33}},\ \bibinfo {pages}
  {115009} (\bibinfo {year} {2016})},\ \Eprint
  {http://arxiv.org/abs/1510.08432} {arXiv:1510.08432 [gr-qc]} \BibitemShut
  {NoStop}%
\bibitem [{\citenamefont {Golovnev}\ \emph {et~al.}(2017)\citenamefont
  {Golovnev}, \citenamefont {Koivisto},\ and\ \citenamefont
  {Sandstad}}]{Golovnev:2017dox}%
  \BibitemOpen
  \bibfield  {author} {\bibinfo {author} {\bibfnamefont {Alexey}\ \bibnamefont
  {Golovnev}}, \bibinfo {author} {\bibfnamefont {Tomi}\ \bibnamefont
  {Koivisto}}, \ and\ \bibinfo {author} {\bibfnamefont {Marit}\ \bibnamefont
  {Sandstad}},\ }\bibfield  {title} {\enquote {\bibinfo {title} {{On the
  covariance of teleparallel gravity theories}},}\ }\href {\doibase
  10.1088/1361-6382/aa7830} {\bibfield  {journal} {\bibinfo  {journal} {Class.
  Quant. Grav.}\ }\textbf {\bibinfo {volume} {34}},\ \bibinfo {pages} {145013}
  (\bibinfo {year} {2017})},\ \Eprint {http://arxiv.org/abs/1701.06271}
  {arXiv:1701.06271 [gr-qc]} \BibitemShut {NoStop}%
\bibitem [{\citenamefont {Krssak}\ \emph {et~al.}(2019)\citenamefont {Krssak},
  \citenamefont {Van Den~Hoogen}, \citenamefont {Pereira}, \citenamefont
  {Boehmer},\ and\ \citenamefont {Coley}}]{Krssak:2018ywd}%
  \BibitemOpen
  \bibfield  {author} {\bibinfo {author} {\bibfnamefont {M.}~\bibnamefont
  {Krssak}}, \bibinfo {author} {\bibfnamefont {R.~J.}\ \bibnamefont {Van
  Den~Hoogen}}, \bibinfo {author} {\bibfnamefont {J.~G.}\ \bibnamefont
  {Pereira}}, \bibinfo {author} {\bibfnamefont {C.~G.}\ \bibnamefont
  {Boehmer}}, \ and\ \bibinfo {author} {\bibfnamefont {A.~A.}\ \bibnamefont
  {Coley}},\ }\bibfield  {title} {\enquote {\bibinfo {title} {{Teleparallel
  Theories of Gravity: Illuminating a Fully Invariant Approach}},}\ }\href
  {\doibase 10.1088/1361-6382/ab2e1f} {\bibfield  {journal} {\bibinfo
  {journal} {Class. Quant. Grav.}\ }\textbf {\bibinfo {volume} {36}},\ \bibinfo
  {pages} {183001} (\bibinfo {year} {2019})},\ \Eprint
  {http://arxiv.org/abs/1810.12932} {arXiv:1810.12932 [gr-qc]} \BibitemShut
  {NoStop}%
\bibitem [{\citenamefont {Bejarano}\ \emph {et~al.}(2019)\citenamefont
  {Bejarano}, \citenamefont {Ferraro}, \citenamefont {Fiorini},\ and\
  \citenamefont {Guzmán}}]{Bejarano:2019fii}%
  \BibitemOpen
  \bibfield  {author} {\bibinfo {author} {\bibfnamefont {Cecilia}\ \bibnamefont
  {Bejarano}}, \bibinfo {author} {\bibfnamefont {Rafael}\ \bibnamefont
  {Ferraro}}, \bibinfo {author} {\bibfnamefont {Franco}\ \bibnamefont
  {Fiorini}}, \ and\ \bibinfo {author} {\bibfnamefont {María~José}\
  \bibnamefont {Guzmán}},\ }\bibfield  {title} {\enquote {\bibinfo {title}
  {{Reflections on the covariance of modified teleparallel theories of
  gravity}},}\ }\href {\doibase 10.3390/universe5060158} {\bibfield  {journal}
  {\bibinfo  {journal} {Universe}\ }\textbf {\bibinfo {volume} {5}},\ \bibinfo
  {pages} {158} (\bibinfo {year} {2019})},\ \Eprint
  {http://arxiv.org/abs/1905.09913} {arXiv:1905.09913 [gr-qc]} \BibitemShut
  {NoStop}%
\bibitem [{\citenamefont {Geng}\ \emph {et~al.}(2011)\citenamefont {Geng},
  \citenamefont {Lee}, \citenamefont {Saridakis},\ and\ \citenamefont
  {Wu}}]{Geng:2011aj}%
  \BibitemOpen
  \bibfield  {author} {\bibinfo {author} {\bibfnamefont {Chao-Qiang}\
  \bibnamefont {Geng}}, \bibinfo {author} {\bibfnamefont {Chung-Chi}\
  \bibnamefont {Lee}}, \bibinfo {author} {\bibfnamefont {Emmanuel~N.}\
  \bibnamefont {Saridakis}}, \ and\ \bibinfo {author} {\bibfnamefont {Yi-Peng}\
  \bibnamefont {Wu}},\ }\bibfield  {title} {\enquote {\bibinfo {title}
  {{“Teleparallel” dark energy}},}\ }\href {\doibase
  10.1016/j.physletb.2011.09.082} {\bibfield  {journal} {\bibinfo  {journal}
  {Phys. Lett.}\ }\textbf {\bibinfo {volume} {B704}},\ \bibinfo {pages}
  {384--387} (\bibinfo {year} {2011})},\ \Eprint
  {http://arxiv.org/abs/1109.1092} {arXiv:1109.1092 [hep-th]} \BibitemShut
  {NoStop}%
\bibitem [{\citenamefont {Chakrabarti}\ \emph {et~al.}(2017)\citenamefont
  {Chakrabarti}, \citenamefont {Said},\ and\ \citenamefont
  {Farrugia}}]{Chakrabarti:2017moe}%
  \BibitemOpen
  \bibfield  {author} {\bibinfo {author} {\bibfnamefont {Soumya}\ \bibnamefont
  {Chakrabarti}}, \bibinfo {author} {\bibfnamefont {Jackson~Levi}\ \bibnamefont
  {Said}}, \ and\ \bibinfo {author} {\bibfnamefont {Gabriel}\ \bibnamefont
  {Farrugia}},\ }\bibfield  {title} {\enquote {\bibinfo {title} {{Some aspects
  of reconstruction using a scalar field in $f(T)$ gravity}},}\ }\href
  {\doibase 10.1140/epjc/s10052-017-5404-6} {\bibfield  {journal} {\bibinfo
  {journal} {Eur. Phys. J.}\ }\textbf {\bibinfo {volume} {C77}},\ \bibinfo
  {pages} {815} (\bibinfo {year} {2017})},\ \Eprint
  {http://arxiv.org/abs/1711.04423} {arXiv:1711.04423 [gr-qc]} \BibitemShut
  {NoStop}%
\bibitem [{\citenamefont {Otalora}(2013)}]{Otalora:2013tba}%
  \BibitemOpen
  \bibfield  {author} {\bibinfo {author} {\bibfnamefont {G.}~\bibnamefont
  {Otalora}},\ }\bibfield  {title} {\enquote {\bibinfo {title} {{Scaling
  attractors in interacting teleparallel dark energy}},}\ }\href {\doibase
  10.1088/1475-7516/2013/07/044} {\bibfield  {journal} {\bibinfo  {journal}
  {JCAP}\ }\textbf {\bibinfo {volume} {1307}},\ \bibinfo {pages} {044}
  (\bibinfo {year} {2013})},\ \Eprint {http://arxiv.org/abs/1305.0474}
  {arXiv:1305.0474 [gr-qc]} \BibitemShut {NoStop}%
\bibitem [{\citenamefont {Jamil}\ \emph {et~al.}(2012)\citenamefont {Jamil},
  \citenamefont {Momeni},\ and\ \citenamefont {Myrzakulov}}]{Jamil:2012vb}%
  \BibitemOpen
  \bibfield  {author} {\bibinfo {author} {\bibfnamefont {Mubasher}\
  \bibnamefont {Jamil}}, \bibinfo {author} {\bibfnamefont {D.}~\bibnamefont
  {Momeni}}, \ and\ \bibinfo {author} {\bibfnamefont {Ratbay}\ \bibnamefont
  {Myrzakulov}},\ }\bibfield  {title} {\enquote {\bibinfo {title} {{Stability
  of a non-minimally conformally coupled scalar field in F(T) cosmology}},}\
  }\href {\doibase 10.1140/epjc/s10052-012-2075-1} {\bibfield  {journal}
  {\bibinfo  {journal} {Eur. Phys. J.}\ }\textbf {\bibinfo {volume} {C72}},\
  \bibinfo {pages} {2075} (\bibinfo {year} {2012})},\ \Eprint
  {http://arxiv.org/abs/1208.0025} {arXiv:1208.0025 [gr-qc]} \BibitemShut
  {NoStop}%
\bibitem [{\citenamefont {Chen}\ \emph
  {et~al.}(2015{\natexlab{b}})\citenamefont {Chen}, \citenamefont {Wu},\ and\
  \citenamefont {Wei}}]{Chen:2014qsa}%
  \BibitemOpen
  \bibfield  {author} {\bibinfo {author} {\bibfnamefont {Zu-Cheng}\
  \bibnamefont {Chen}}, \bibinfo {author} {\bibfnamefont {You}\ \bibnamefont
  {Wu}}, \ and\ \bibinfo {author} {\bibfnamefont {Hao}\ \bibnamefont {Wei}},\
  }\bibfield  {title} {\enquote {\bibinfo {title} {{Post-Newtonian
  Approximation of Teleparallel Gravity Coupled with a Scalar Field}},}\ }\href
  {\doibase 10.1016/j.nuclphysb.2015.03.012} {\bibfield  {journal} {\bibinfo
  {journal} {Nucl. Phys.}\ }\textbf {\bibinfo {volume} {B894}},\ \bibinfo
  {pages} {422--438} (\bibinfo {year} {2015}{\natexlab{b}})},\ \Eprint
  {http://arxiv.org/abs/1410.7715} {arXiv:1410.7715 [gr-qc]} \BibitemShut
  {NoStop}%
\bibitem [{\citenamefont {Bahamonde}\ and\ \citenamefont
  {Wright}(2015)}]{Bahamonde:2015hza}%
  \BibitemOpen
  \bibfield  {author} {\bibinfo {author} {\bibfnamefont {Sebastian}\
  \bibnamefont {Bahamonde}}\ and\ \bibinfo {author} {\bibfnamefont {Matthew}\
  \bibnamefont {Wright}},\ }\bibfield  {title} {\enquote {\bibinfo {title}
  {{Teleparallel quintessence with a nonminimal coupling to a boundary
  term}},}\ }\href {\doibase 10.1103/PhysRevD.92.084034,
  10.1103/PhysRevD.93.109901} {\bibfield  {journal} {\bibinfo  {journal} {Phys.
  Rev.}\ }\textbf {\bibinfo {volume} {D92}},\ \bibinfo {pages} {084034}
  (\bibinfo {year} {2015})},\ \bibinfo {note} {[Erratum: Phys.
  Rev.D93,no.10,109901(2016)]},\ \Eprint {http://arxiv.org/abs/1508.06580}
  {arXiv:1508.06580 [gr-qc]} \BibitemShut {NoStop}%
\bibitem [{\citenamefont {Bamba}\ \emph {et~al.}(2013)\citenamefont {Bamba},
  \citenamefont {Odintsov},\ and\ \citenamefont
  {Sáez-Gómez}}]{Bamba:2013jqa}%
  \BibitemOpen
  \bibfield  {author} {\bibinfo {author} {\bibfnamefont {Kazuharu}\
  \bibnamefont {Bamba}}, \bibinfo {author} {\bibfnamefont {Sergei~D.}\
  \bibnamefont {Odintsov}}, \ and\ \bibinfo {author} {\bibfnamefont {Diego}\
  \bibnamefont {Sáez-Gómez}},\ }\bibfield  {title} {\enquote {\bibinfo
  {title} {{Conformal symmetry and accelerating cosmology in teleparallel
  gravity}},}\ }\href {\doibase 10.1103/PhysRevD.88.084042} {\bibfield
  {journal} {\bibinfo  {journal} {Phys. Rev.}\ }\textbf {\bibinfo {volume}
  {D88}},\ \bibinfo {pages} {084042} (\bibinfo {year} {2013})},\ \Eprint
  {http://arxiv.org/abs/1308.5789} {arXiv:1308.5789 [gr-qc]} \BibitemShut
  {NoStop}%
\bibitem [{\citenamefont {Nojiri}\ \emph {et~al.}(2017)\citenamefont {Nojiri},
  \citenamefont {Odintsov},\ and\ \citenamefont {Oikonomou}}]{Nojiri:2017ncd}%
  \BibitemOpen
  \bibfield  {author} {\bibinfo {author} {\bibfnamefont {S.}~\bibnamefont
  {Nojiri}}, \bibinfo {author} {\bibfnamefont {S.~D.}\ \bibnamefont
  {Odintsov}}, \ and\ \bibinfo {author} {\bibfnamefont {V.~K.}\ \bibnamefont
  {Oikonomou}},\ }\bibfield  {title} {\enquote {\bibinfo {title} {{Modified
  Gravity Theories on a Nutshell: Inflation, Bounce and Late-time
  Evolution}},}\ }\href {\doibase 10.1016/j.physrep.2017.06.001} {\bibfield
  {journal} {\bibinfo  {journal} {Phys. Rept.}\ }\textbf {\bibinfo {volume}
  {692}},\ \bibinfo {pages} {1--104} (\bibinfo {year} {2017})},\ \Eprint
  {http://arxiv.org/abs/1705.11098} {arXiv:1705.11098 [gr-qc]} \BibitemShut
  {NoStop}%
\bibitem [{\citenamefont {Hohmann}\ \emph {et~al.}(2018)\citenamefont
  {Hohmann}, \citenamefont {Järv},\ and\ \citenamefont
  {Ualikhanova}}]{Hohmann:2018rwf}%
  \BibitemOpen
  \bibfield  {author} {\bibinfo {author} {\bibfnamefont {Manuel}\ \bibnamefont
  {Hohmann}}, \bibinfo {author} {\bibfnamefont {Laur}\ \bibnamefont {Järv}}, \
  and\ \bibinfo {author} {\bibfnamefont {Ulbossyn}\ \bibnamefont
  {Ualikhanova}},\ }\bibfield  {title} {\enquote {\bibinfo {title} {{Covariant
  formulation of scalar-torsion gravity}},}\ }\href {\doibase
  10.1103/PhysRevD.97.104011} {\bibfield  {journal} {\bibinfo  {journal} {Phys.
  Rev.}\ }\textbf {\bibinfo {volume} {D97}},\ \bibinfo {pages} {104011}
  (\bibinfo {year} {2018})},\ \Eprint {http://arxiv.org/abs/1801.05786}
  {arXiv:1801.05786 [gr-qc]} \BibitemShut {NoStop}%
\bibitem [{\citenamefont {Hohmann}(2018{\natexlab{a}})}]{Hohmann:2018vle}%
  \BibitemOpen
  \bibfield  {author} {\bibinfo {author} {\bibfnamefont {Manuel}\ \bibnamefont
  {Hohmann}},\ }\bibfield  {title} {\enquote {\bibinfo {title} {{Scalar-torsion
  theories of gravity I: general formalism and conformal transformations}},}\
  }\href {\doibase 10.1103/PhysRevD.98.064002} {\bibfield  {journal} {\bibinfo
  {journal} {Phys. Rev.}\ }\textbf {\bibinfo {volume} {D98}},\ \bibinfo {pages}
  {064002} (\bibinfo {year} {2018}{\natexlab{a}})},\ \Eprint
  {http://arxiv.org/abs/1801.06528} {arXiv:1801.06528 [gr-qc]} \BibitemShut
  {NoStop}%
\bibitem [{\citenamefont {Hohmann}\ and\ \citenamefont
  {Pfeifer}(2018)}]{Hohmann:2018dqh}%
  \BibitemOpen
  \bibfield  {author} {\bibinfo {author} {\bibfnamefont {Manuel}\ \bibnamefont
  {Hohmann}}\ and\ \bibinfo {author} {\bibfnamefont {Christian}\ \bibnamefont
  {Pfeifer}},\ }\bibfield  {title} {\enquote {\bibinfo {title} {{Scalar-torsion
  theories of gravity II: $L(T, X, Y, \phi)$ theory}},}\ }\href {\doibase
  10.1103/PhysRevD.98.064003} {\bibfield  {journal} {\bibinfo  {journal} {Phys.
  Rev.}\ }\textbf {\bibinfo {volume} {D98}},\ \bibinfo {pages} {064003}
  (\bibinfo {year} {2018})},\ \Eprint {http://arxiv.org/abs/1801.06536}
  {arXiv:1801.06536 [gr-qc]} \BibitemShut {NoStop}%
\bibitem [{\citenamefont {Hohmann}(2018{\natexlab{b}})}]{Hohmann:2018ijr}%
  \BibitemOpen
  \bibfield  {author} {\bibinfo {author} {\bibfnamefont {Manuel}\ \bibnamefont
  {Hohmann}},\ }\bibfield  {title} {\enquote {\bibinfo {title} {{Scalar-torsion
  theories of gravity III: analogue of scalar-tensor gravity and conformal
  invariants}},}\ }\href {\doibase 10.1103/PhysRevD.98.064004} {\bibfield
  {journal} {\bibinfo  {journal} {Phys. Rev.}\ }\textbf {\bibinfo {volume}
  {D98}},\ \bibinfo {pages} {064004} (\bibinfo {year} {2018}{\natexlab{b}})},\
  \Eprint {http://arxiv.org/abs/1801.06531} {arXiv:1801.06531 [gr-qc]}
  \BibitemShut {NoStop}%
\bibitem [{\citenamefont {Will}(1993)}]{Will:1993ns}%
  \BibitemOpen
  \bibfield  {author} {\bibinfo {author} {\bibfnamefont {Clifford.~M.}\
  \bibnamefont {Will}},\ }\href {\doibase 10.1017/CBO9780511564246} {\emph
  {\bibinfo {title} {{Theory and experiment in gravitational physics}}}}\
  (\bibinfo  {publisher} {Cambridge University Press},\ \bibinfo {year}
  {1993})\BibitemShut {NoStop}%
\bibitem [{\citenamefont {Will}(2014)}]{Will:2014kxa}%
  \BibitemOpen
  \bibfield  {author} {\bibinfo {author} {\bibfnamefont {Clifford~M.}\
  \bibnamefont {Will}},\ }\bibfield  {title} {\enquote {\bibinfo {title} {{The
  Confrontation between General Relativity and Experiment}},}\ }\href {\doibase
  10.12942/lrr-2014-4} {\bibfield  {journal} {\bibinfo  {journal} {Living Rev.
  Rel.}\ }\textbf {\bibinfo {volume} {17}},\ \bibinfo {pages} {4} (\bibinfo
  {year} {2014})},\ \Eprint {http://arxiv.org/abs/1403.7377} {arXiv:1403.7377
  [gr-qc]} \BibitemShut {NoStop}%
\bibitem [{\citenamefont {Will}(2018)}]{Will:2018bme}%
  \BibitemOpen
  \bibfield  {author} {\bibinfo {author} {\bibfnamefont {Clifford~M.}\
  \bibnamefont {Will}},\ }\href
  {https://www.cambridge.org/academic/subjects/physics/cosmology-relativity-and-gravitation/theory-and-experiment-gravitational-physics-2nd-edition?format=AR&isbn=9781108679824}
  {\emph {\bibinfo {title} {{Theory and Experiment in Gravitational
  Physics}}}}\ (\bibinfo  {publisher} {Cambridge University Press},\ \bibinfo
  {year} {2018})\BibitemShut {NoStop}%
\bibitem [{\citenamefont {Li}\ \emph {et~al.}(2014)\citenamefont {Li},
  \citenamefont {Wu},\ and\ \citenamefont {Geng}}]{Li:2013oef}%
  \BibitemOpen
  \bibfield  {author} {\bibinfo {author} {\bibfnamefont {Jung-Tsung}\
  \bibnamefont {Li}}, \bibinfo {author} {\bibfnamefont {Yi-Peng}\ \bibnamefont
  {Wu}}, \ and\ \bibinfo {author} {\bibfnamefont {Chao-Qiang}\ \bibnamefont
  {Geng}},\ }\bibfield  {title} {\enquote {\bibinfo {title} {{Parametrized
  post-Newtonian limit of the teleparallel dark energy model}},}\ }\href
  {\doibase 10.1103/PhysRevD.89.044040} {\bibfield  {journal} {\bibinfo
  {journal} {Phys. Rev.}\ }\textbf {\bibinfo {volume} {D89}},\ \bibinfo {pages}
  {044040} (\bibinfo {year} {2014})},\ \Eprint {http://arxiv.org/abs/1312.4332}
  {arXiv:1312.4332 [gr-qc]} \BibitemShut {NoStop}%
\bibitem [{\citenamefont {Mohseni~Sadjadi}(2017)}]{Sadjadi:2016kwj}%
  \BibitemOpen
  \bibfield  {author} {\bibinfo {author} {\bibfnamefont {H.}~\bibnamefont
  {Mohseni~Sadjadi}},\ }\bibfield  {title} {\enquote {\bibinfo {title}
  {{Parameterized post-Newtonian approximation in a teleparallel model of dark
  energy with a boundary term}},}\ }\href {\doibase
  10.1140/epjc/s10052-017-4760-6} {\bibfield  {journal} {\bibinfo  {journal}
  {Eur. Phys. J.}\ }\textbf {\bibinfo {volume} {C77}},\ \bibinfo {pages} {191}
  (\bibinfo {year} {2017})},\ \Eprint {http://arxiv.org/abs/1606.04362}
  {arXiv:1606.04362 [gr-qc]} \BibitemShut {NoStop}%
\bibitem [{\citenamefont {Ualikhanova}\ and\ \citenamefont
  {Hohmann}(2019)}]{Ualikhanova:2019ygl}%
  \BibitemOpen
  \bibfield  {author} {\bibinfo {author} {\bibfnamefont {Ulbossyn}\
  \bibnamefont {Ualikhanova}}\ and\ \bibinfo {author} {\bibfnamefont {Manuel}\
  \bibnamefont {Hohmann}},\ }\bibfield  {title} {\enquote {\bibinfo {title}
  {{Parameterized post-Newtonian limit of general teleparallel gravity
  theories}},}\ }\href {\doibase 10.1103/PhysRevD.100.104011} {\bibfield
  {journal} {\bibinfo  {journal} {Phys. Rev.}\ }\textbf {\bibinfo {volume}
  {D100}},\ \bibinfo {pages} {104011} (\bibinfo {year} {2019})},\ \Eprint
  {http://arxiv.org/abs/1907.08178} {arXiv:1907.08178 [gr-qc]} \BibitemShut
  {NoStop}%
\bibitem [{\citenamefont {Hayward}(1981)}]{Hayward:1981bk}%
  \BibitemOpen
  \bibfield  {author} {\bibinfo {author} {\bibfnamefont {J.}~\bibnamefont
  {Hayward}},\ }\bibfield  {title} {\enquote {\bibinfo {title} {{Scalar tetrad
  theories of gravity}},}\ }\href {\doibase 10.1007/BF00766297} {\bibfield
  {journal} {\bibinfo  {journal} {Gen. Rel. Grav.}\ }\textbf {\bibinfo {volume}
  {13}},\ \bibinfo {pages} {43--55} (\bibinfo {year} {1981})}\BibitemShut
  {NoStop}%
\bibitem [{\citenamefont {Schweizer}\ and\ \citenamefont
  {Straumann}(1979)}]{Schweizer:1979up}%
  \BibitemOpen
  \bibfield  {author} {\bibinfo {author} {\bibfnamefont {M.}~\bibnamefont
  {Schweizer}}\ and\ \bibinfo {author} {\bibfnamefont {N.}~\bibnamefont
  {Straumann}},\ }\bibfield  {title} {\enquote {\bibinfo {title} {{Poincare
  gauge theory of gravitation and the binary pulsar 1913+16}},}\ }\href
  {\doibase 10.1016/0375-9601(79)90645-5} {\bibfield  {journal} {\bibinfo
  {journal} {Phys. Lett.}\ }\textbf {\bibinfo {volume} {71A}},\ \bibinfo
  {pages} {493--495} (\bibinfo {year} {1979})}\BibitemShut {NoStop}%
\bibitem [{\citenamefont {Schweizer}\ \emph {et~al.}(1980)\citenamefont
  {Schweizer}, \citenamefont {Straumann},\ and\ \citenamefont
  {Wipf}}]{Schweizer:1980vn}%
  \BibitemOpen
  \bibfield  {author} {\bibinfo {author} {\bibfnamefont {M.}~\bibnamefont
  {Schweizer}}, \bibinfo {author} {\bibfnamefont {N.}~\bibnamefont
  {Straumann}}, \ and\ \bibinfo {author} {\bibfnamefont {A.}~\bibnamefont
  {Wipf}},\ }\bibfield  {title} {\enquote {\bibinfo {title} {{Postnewtonian
  generation of gravitational waves in a theory of gravity with torsion}},}\
  }\href {\doibase 10.1007/BF00757366} {\bibfield  {journal} {\bibinfo
  {journal} {Gen. Rel. Grav.}\ }\textbf {\bibinfo {volume} {12}},\ \bibinfo
  {pages} {951--961} (\bibinfo {year} {1980})}\BibitemShut {NoStop}%
\bibitem [{\citenamefont {Smalley}(1980)}]{Smalley:1980em}%
  \BibitemOpen
  \bibfield  {author} {\bibinfo {author} {\bibfnamefont {L.~L.}\ \bibnamefont
  {Smalley}},\ }\bibfield  {title} {\enquote {\bibinfo {title} {{Postnewtonian
  approximation of the Poincare gauge theory of gravitation}},}\ }\href
  {\doibase 10.1103/PhysRevD.21.328} {\bibfield  {journal} {\bibinfo  {journal}
  {Phys. Rev.}\ }\textbf {\bibinfo {volume} {D21}},\ \bibinfo {pages}
  {328--331} (\bibinfo {year} {1980})}\BibitemShut {NoStop}%
\bibitem [{\citenamefont {Nitsch}\ and\ \citenamefont
  {Hehl}(1980)}]{Nitsch:1979qn}%
  \BibitemOpen
  \bibfield  {author} {\bibinfo {author} {\bibfnamefont {Jurgen}\ \bibnamefont
  {Nitsch}}\ and\ \bibinfo {author} {\bibfnamefont {Friedrich~W.}\ \bibnamefont
  {Hehl}},\ }\bibfield  {title} {\enquote {\bibinfo {title} {{Translational
  Gauge Theory of Gravity: Postnewtonian Approximation and Spin Precession}},}\
  }\href {\doibase 10.1016/0370-2693(80)90059-3} {\bibfield  {journal}
  {\bibinfo  {journal} {Phys. Lett.}\ }\textbf {\bibinfo {volume} {90B}},\
  \bibinfo {pages} {98--102} (\bibinfo {year} {1980})}\BibitemShut {NoStop}%
\bibitem [{\citenamefont {Gladchenko}\ \emph {et~al.}(1990)\citenamefont
  {Gladchenko}, \citenamefont {Ponomarev},\ and\ \citenamefont
  {Zhytnikov}}]{Gladchenko:1990nw}%
  \BibitemOpen
  \bibfield  {author} {\bibinfo {author} {\bibfnamefont {M.~S.}\ \bibnamefont
  {Gladchenko}}, \bibinfo {author} {\bibfnamefont {V.~N.}\ \bibnamefont
  {Ponomarev}}, \ and\ \bibinfo {author} {\bibfnamefont {V.~V.}\ \bibnamefont
  {Zhytnikov}},\ }\bibfield  {title} {\enquote {\bibinfo {title} {{PPN metric
  and PPN torsion in the quadratic Poincare gauge theory of gravity}},}\ }\href
  {\doibase 10.1016/0370-2693(90)91488-W} {\bibfield  {journal} {\bibinfo
  {journal} {Phys. Lett.}\ }\textbf {\bibinfo {volume} {B241}},\ \bibinfo
  {pages} {67--69} (\bibinfo {year} {1990})}\BibitemShut {NoStop}%
\bibitem [{\citenamefont {Gladchenko}\ and\ \citenamefont
  {Zhytnikov}(1994)}]{Gladchenko:1994wu}%
  \BibitemOpen
  \bibfield  {author} {\bibinfo {author} {\bibfnamefont {M.~S.}\ \bibnamefont
  {Gladchenko}}\ and\ \bibinfo {author} {\bibfnamefont {V.~V.}\ \bibnamefont
  {Zhytnikov}},\ }\bibfield  {title} {\enquote {\bibinfo {title}
  {{PostNewtonian effects in the quadratic Poincare gauge theory of
  gravitation}},}\ }\href {\doibase 10.1103/PhysRevD.50.5060} {\bibfield
  {journal} {\bibinfo  {journal} {Phys. Rev.}\ }\textbf {\bibinfo {volume}
  {D50}},\ \bibinfo {pages} {5060--5071} (\bibinfo {year} {1994})}\BibitemShut
  {NoStop}%
\bibitem [{\citenamefont {Jarv}\ and\ \citenamefont
  {Toporensky}(2016)}]{Jarv:2015odu}%
  \BibitemOpen
  \bibfield  {author} {\bibinfo {author} {\bibfnamefont {Laur}\ \bibnamefont
  {Jarv}}\ and\ \bibinfo {author} {\bibfnamefont {Alexey}\ \bibnamefont
  {Toporensky}},\ }\bibfield  {title} {\enquote {\bibinfo {title} {{General
  relativity as an attractor for scalar-torsion cosmology}},}\ }\href {\doibase
  10.1103/PhysRevD.93.024051} {\bibfield  {journal} {\bibinfo  {journal} {Phys.
  Rev.}\ }\textbf {\bibinfo {volume} {D93}},\ \bibinfo {pages} {024051}
  (\bibinfo {year} {2016})},\ \Eprint {http://arxiv.org/abs/1511.03933}
  {arXiv:1511.03933 [gr-qc]} \BibitemShut {NoStop}%
\bibitem [{\citenamefont {Järv}(2017)}]{Jarv:2017npl}%
  \BibitemOpen
  \bibfield  {author} {\bibinfo {author} {\bibfnamefont {Laur}\ \bibnamefont
  {Järv}},\ }\bibfield  {title} {\enquote {\bibinfo {title} {{Effective
  Gravitational “Constant” in Scalar-(Curvature)Tensor and Scalar-Torsion
  Gravities}},}\ }\href {\doibase 10.3390/universe3020037} {\bibfield
  {journal} {\bibinfo  {journal} {Universe}\ }\textbf {\bibinfo {volume} {3}},\
  \bibinfo {pages} {37} (\bibinfo {year} {2017})}\BibitemShut {NoStop}%
\bibitem [{\citenamefont {Olmo}(2005{\natexlab{a}})}]{Olmo:2005zr}%
  \BibitemOpen
  \bibfield  {author} {\bibinfo {author} {\bibfnamefont {Gonzalo~J.}\
  \bibnamefont {Olmo}},\ }\bibfield  {title} {\enquote {\bibinfo {title} {{The
  Gravity Lagrangian according to solar system experiments}},}\ }\href
  {\doibase 10.1103/PhysRevLett.95.261102} {\bibfield  {journal} {\bibinfo
  {journal} {Phys. Rev. Lett.}\ }\textbf {\bibinfo {volume} {95}},\ \bibinfo
  {pages} {261102} (\bibinfo {year} {2005}{\natexlab{a}})},\ \Eprint
  {http://arxiv.org/abs/gr-qc/0505101} {arXiv:gr-qc/0505101 [gr-qc]}
  \BibitemShut {NoStop}%
\bibitem [{\citenamefont {Olmo}(2005{\natexlab{b}})}]{Olmo:2005hc}%
  \BibitemOpen
  \bibfield  {author} {\bibinfo {author} {\bibfnamefont {Gonzalo~J.}\
  \bibnamefont {Olmo}},\ }\bibfield  {title} {\enquote {\bibinfo {title}
  {{Post-Newtonian constraints on f(R) cosmologies in metric and Palatini
  formalism}},}\ }\href {\doibase 10.1103/PhysRevD.72.083505} {\bibfield
  {journal} {\bibinfo  {journal} {Phys. Rev.}\ }\textbf {\bibinfo {volume}
  {D72}},\ \bibinfo {pages} {083505} (\bibinfo {year} {2005}{\natexlab{b}})},\
  \Eprint {http://arxiv.org/abs/gr-qc/0505135} {arXiv:gr-qc/0505135 [gr-qc]}
  \BibitemShut {NoStop}%
\bibitem [{\citenamefont {Perivolaropoulos}(2010)}]{Perivolaropoulos:2009ak}%
  \BibitemOpen
  \bibfield  {author} {\bibinfo {author} {\bibfnamefont {L.}~\bibnamefont
  {Perivolaropoulos}},\ }\bibfield  {title} {\enquote {\bibinfo {title} {{PPN
  Parameter gamma and Solar System Constraints of Massive Brans-Dicke
  Theories}},}\ }\href {\doibase 10.1103/PhysRevD.81.047501} {\bibfield
  {journal} {\bibinfo  {journal} {Phys. Rev.}\ }\textbf {\bibinfo {volume}
  {D81}},\ \bibinfo {pages} {047501} (\bibinfo {year} {2010})},\ \Eprint
  {http://arxiv.org/abs/0911.3401} {arXiv:0911.3401 [gr-qc]} \BibitemShut
  {NoStop}%
\bibitem [{\citenamefont {Hohmann}\ \emph {et~al.}(2013)\citenamefont
  {Hohmann}, \citenamefont {Jarv}, \citenamefont {Kuusk},\ and\ \citenamefont
  {Randla}}]{Hohmann:2013rba}%
  \BibitemOpen
  \bibfield  {author} {\bibinfo {author} {\bibfnamefont {Manuel}\ \bibnamefont
  {Hohmann}}, \bibinfo {author} {\bibfnamefont {Laur}\ \bibnamefont {Jarv}},
  \bibinfo {author} {\bibfnamefont {Piret}\ \bibnamefont {Kuusk}}, \ and\
  \bibinfo {author} {\bibfnamefont {Erik}\ \bibnamefont {Randla}},\ }\bibfield
  {title} {\enquote {\bibinfo {title} {{Post-Newtonian parameters $\gamma$ and
  $\beta$ of scalar-tensor gravity with a general potential}},}\ }\href
  {\doibase 10.1103/PhysRevD.89.069901, 10.1103/PhysRevD.88.084054} {\bibfield
  {journal} {\bibinfo  {journal} {Phys. Rev.}\ }\textbf {\bibinfo {volume}
  {D88}},\ \bibinfo {pages} {084054} (\bibinfo {year} {2013})},\ \bibinfo
  {note} {[Erratum: Phys. Rev.D89,no.6,069901(2014)]},\ \Eprint
  {http://arxiv.org/abs/1309.0031} {arXiv:1309.0031 [gr-qc]} \BibitemShut
  {NoStop}%
\bibitem [{\citenamefont {Schärer}\ \emph {et~al.}(2014)\citenamefont
  {Schärer}, \citenamefont {Angélil}, \citenamefont {Bondarescu},
  \citenamefont {Jetzer},\ and\ \citenamefont {Lundgren}}]{Scharer:2014kya}%
  \BibitemOpen
  \bibfield  {author} {\bibinfo {author} {\bibfnamefont {Andreas}\ \bibnamefont
  {Schärer}}, \bibinfo {author} {\bibfnamefont {Raymond}\ \bibnamefont
  {Angélil}}, \bibinfo {author} {\bibfnamefont {Ruxandra}\ \bibnamefont
  {Bondarescu}}, \bibinfo {author} {\bibfnamefont {Philippe}\ \bibnamefont
  {Jetzer}}, \ and\ \bibinfo {author} {\bibfnamefont {Andrew}\ \bibnamefont
  {Lundgren}},\ }\bibfield  {title} {\enquote {\bibinfo {title} {{Testing
  scalar-tensor theories and parametrized post-Newtonian parameters in Earth
  orbit}},}\ }\href {\doibase 10.1103/PhysRevD.90.123005} {\bibfield  {journal}
  {\bibinfo  {journal} {Phys. Rev.}\ }\textbf {\bibinfo {volume} {D90}},\
  \bibinfo {pages} {123005} (\bibinfo {year} {2014})},\ \Eprint
  {http://arxiv.org/abs/1410.7914} {arXiv:1410.7914 [gr-qc]} \BibitemShut
  {NoStop}%
\bibitem [{\citenamefont {Hohmann}(2015)}]{Hohmann:2015kra}%
  \BibitemOpen
  \bibfield  {author} {\bibinfo {author} {\bibfnamefont {Manuel}\ \bibnamefont
  {Hohmann}},\ }\bibfield  {title} {\enquote {\bibinfo {title} {{Parametrized
  post-Newtonian limit of Horndeski’s gravity theory}},}\ }\href {\doibase
  10.1103/PhysRevD.92.064019} {\bibfield  {journal} {\bibinfo  {journal} {Phys.
  Rev.}\ }\textbf {\bibinfo {volume} {D92}},\ \bibinfo {pages} {064019}
  (\bibinfo {year} {2015})},\ \Eprint {http://arxiv.org/abs/1506.04253}
  {arXiv:1506.04253 [gr-qc]} \BibitemShut {NoStop}%
\bibitem [{\citenamefont {Hohmann}(2019{\natexlab{a}})}]{Hohmann:2019qgo}%
  \BibitemOpen
  \bibfield  {author} {\bibinfo {author} {\bibfnamefont {Manuel}\ \bibnamefont
  {Hohmann}},\ }\bibfield  {title} {\enquote {\bibinfo {title}
  {{Gauge-invariant approach to the parametrized post-Newtonian formalism}},}\
  }\href@noop {} {\  (\bibinfo {year} {2019}{\natexlab{a}})},\ \Eprint
  {http://arxiv.org/abs/1910.09245} {arXiv:1910.09245 [gr-qc]} \BibitemShut
  {NoStop}%
\bibitem [{\citenamefont {Nutku}(1969)}]{Nutku:1969}%
  \BibitemOpen
  \bibfield  {author} {\bibinfo {author} {\bibfnamefont {Yavuz}\ \bibnamefont
  {Nutku}},\ }\bibfield  {title} {\enquote {\bibinfo {title} {{The
  Post-Newtonian Equations of Hydrodynamics in the Brans-Dicke Theory}},}\
  }\href {\doibase 10.1086/149928} {\bibfield  {journal} {\bibinfo  {journal}
  {Astrophys. J.}\ }\textbf {\bibinfo {volume} {155}},\ \bibinfo {pages} {999}
  (\bibinfo {year} {1969})}\BibitemShut {NoStop}%
\bibitem [{\citenamefont {Nordtvedt}(1970)}]{Nordtvedt:1970uv}%
  \BibitemOpen
  \bibfield  {author} {\bibinfo {author} {\bibfnamefont {Kenneth}\ \bibnamefont
  {Nordtvedt}, \bibfnamefont {Jr.}},\ }\bibfield  {title} {\enquote {\bibinfo
  {title} {{Post-Newtonian metric for a general class of scalar tensor
  gravitational theories and observational consequences}},}\ }\href {\doibase
  10.1086/150607} {\bibfield  {journal} {\bibinfo  {journal} {Astrophys. J.}\
  }\textbf {\bibinfo {volume} {161}},\ \bibinfo {pages} {1059--1067} (\bibinfo
  {year} {1970})}\BibitemShut {NoStop}%
\bibitem [{\citenamefont {Flathmann}\ and\ \citenamefont
  {Hohmann}(2020)}]{Flathmann:2019khc}%
  \BibitemOpen
  \bibfield  {author} {\bibinfo {author} {\bibfnamefont {Kai}\ \bibnamefont
  {Flathmann}}\ and\ \bibinfo {author} {\bibfnamefont {Manuel}\ \bibnamefont
  {Hohmann}},\ }\bibfield  {title} {\enquote {\bibinfo {title} {{Post-Newtonian
  Limit of Generalized Scalar-Torsion Theories of Gravity}},}\ }\href {\doibase
  10.1103/PhysRevD.101.024005} {\bibfield  {journal} {\bibinfo  {journal}
  {Phys. Rev.}\ }\textbf {\bibinfo {volume} {D101}},\ \bibinfo {pages} {024005}
  (\bibinfo {year} {2020})},\ \Eprint {http://arxiv.org/abs/1910.01023}
  {arXiv:1910.01023 [gr-qc]} \BibitemShut {NoStop}%
\bibitem [{\citenamefont {Bahamonde}\ \emph {et~al.}(2019)\citenamefont
  {Bahamonde}, \citenamefont {Dialektopoulos},\ and\ \citenamefont
  {Said}}]{Bahamonde:2019shr}%
  \BibitemOpen
  \bibfield  {author} {\bibinfo {author} {\bibfnamefont {Sebastian}\
  \bibnamefont {Bahamonde}}, \bibinfo {author} {\bibfnamefont
  {Konstantinos~F.}\ \bibnamefont {Dialektopoulos}}, \ and\ \bibinfo {author}
  {\bibfnamefont {Jackson~Levi}\ \bibnamefont {Said}},\ }\bibfield  {title}
  {\enquote {\bibinfo {title} {{Can Horndeski Theory be recast using
  Teleparallel Gravity?}}}\ }\href {\doibase 10.1103/PhysRevD.100.064018}
  {\bibfield  {journal} {\bibinfo  {journal} {Phys. Rev.}\ }\textbf {\bibinfo
  {volume} {D100}},\ \bibinfo {pages} {064018} (\bibinfo {year} {2019})},\
  \Eprint {http://arxiv.org/abs/1904.10791} {arXiv:1904.10791 [gr-qc]}
  \BibitemShut {NoStop}%
\bibitem [{\citenamefont {Hohmann}(2019{\natexlab{b}})}]{Hohmann:2019gmt}%
  \BibitemOpen
  \bibfield  {author} {\bibinfo {author} {\bibfnamefont {Manuel}\ \bibnamefont
  {Hohmann}},\ }\bibfield  {title} {\enquote {\bibinfo {title} {{Disformal
  Transformations in Scalar-Torsion Gravity}},}\ }\href {\doibase
  10.3390/universe5070167} {\bibfield  {journal} {\bibinfo  {journal}
  {Universe}\ }\textbf {\bibinfo {volume} {5}},\ \bibinfo {pages} {167}
  (\bibinfo {year} {2019}{\natexlab{b}})},\ \Eprint
  {http://arxiv.org/abs/1905.00451} {arXiv:1905.00451 [gr-qc]} \BibitemShut
  {NoStop}%
\bibitem [{\citenamefont {Hohmann}\ \emph {et~al.}(2016)\citenamefont
  {Hohmann}, \citenamefont {Jarv}, \citenamefont {Kuusk}, \citenamefont
  {Randla},\ and\ \citenamefont {Vilson}}]{Hohmann:2016yfd}%
  \BibitemOpen
  \bibfield  {author} {\bibinfo {author} {\bibfnamefont {Manuel}\ \bibnamefont
  {Hohmann}}, \bibinfo {author} {\bibfnamefont {Laur}\ \bibnamefont {Jarv}},
  \bibinfo {author} {\bibfnamefont {Piret}\ \bibnamefont {Kuusk}}, \bibinfo
  {author} {\bibfnamefont {Erik}\ \bibnamefont {Randla}}, \ and\ \bibinfo
  {author} {\bibfnamefont {Ott}\ \bibnamefont {Vilson}},\ }\bibfield  {title}
  {\enquote {\bibinfo {title} {{Post-Newtonian parameter $\gamma$ for
  multiscalar-tensor gravity with a general potential}},}\ }\href {\doibase
  10.1103/PhysRevD.94.124015} {\bibfield  {journal} {\bibinfo  {journal} {Phys.
  Rev.}\ }\textbf {\bibinfo {volume} {D94}},\ \bibinfo {pages} {124015}
  (\bibinfo {year} {2016})},\ \Eprint {http://arxiv.org/abs/1607.02356}
  {arXiv:1607.02356 [gr-qc]} \BibitemShut {NoStop}%
\bibitem [{\citenamefont {Järv}\ \emph {et~al.}(2018)\citenamefont {Järv},
  \citenamefont {Rünkla}, \citenamefont {Saal},\ and\ \citenamefont
  {Vilson}}]{Jarv:2018bgs}%
  \BibitemOpen
  \bibfield  {author} {\bibinfo {author} {\bibfnamefont {Laur}\ \bibnamefont
  {Järv}}, \bibinfo {author} {\bibfnamefont {Mihkel}\ \bibnamefont {Rünkla}},
  \bibinfo {author} {\bibfnamefont {Margus}\ \bibnamefont {Saal}}, \ and\
  \bibinfo {author} {\bibfnamefont {Ott}\ \bibnamefont {Vilson}},\ }\bibfield
  {title} {\enquote {\bibinfo {title} {{Nonmetricity formulation of general
  relativity and its scalar-tensor extension}},}\ }\href {\doibase
  10.1103/PhysRevD.97.124025} {\bibfield  {journal} {\bibinfo  {journal} {Phys.
  Rev.}\ }\textbf {\bibinfo {volume} {D97}},\ \bibinfo {pages} {124025}
  (\bibinfo {year} {2018})},\ \Eprint {http://arxiv.org/abs/1802.00492}
  {arXiv:1802.00492 [gr-qc]} \BibitemShut {NoStop}%
\bibitem [{\citenamefont {Rünkla}\ and\ \citenamefont
  {Vilson}(2018)}]{Runkla:2018xrv}%
  \BibitemOpen
  \bibfield  {author} {\bibinfo {author} {\bibfnamefont {Mihkel}\ \bibnamefont
  {Rünkla}}\ and\ \bibinfo {author} {\bibfnamefont {Ott}\ \bibnamefont
  {Vilson}},\ }\bibfield  {title} {\enquote {\bibinfo {title} {{Family of
  scalar-nonmetricity theories of gravity}},}\ }\href {\doibase
  10.1103/PhysRevD.98.084034} {\bibfield  {journal} {\bibinfo  {journal} {Phys.
  Rev.}\ }\textbf {\bibinfo {volume} {D98}},\ \bibinfo {pages} {084034}
  (\bibinfo {year} {2018})},\ \Eprint {http://arxiv.org/abs/1805.12197}
  {arXiv:1805.12197 [gr-qc]} \BibitemShut {NoStop}%
\bibitem [{\citenamefont {Hohmann}\ and\ \citenamefont
  {Schärer}(2017)}]{Hohmann:2017qje}%
  \BibitemOpen
  \bibfield  {author} {\bibinfo {author} {\bibfnamefont {Manuel}\ \bibnamefont
  {Hohmann}}\ and\ \bibinfo {author} {\bibfnamefont {Andreas}\ \bibnamefont
  {Schärer}},\ }\bibfield  {title} {\enquote {\bibinfo {title}
  {{Post-Newtonian parameters $\gamma$ and $\beta$ of scalar-tensor gravity for
  a homogeneous gravitating sphere}},}\ }\href {\doibase
  10.1103/PhysRevD.96.104026} {\bibfield  {journal} {\bibinfo  {journal} {Phys.
  Rev.}\ }\textbf {\bibinfo {volume} {D96}},\ \bibinfo {pages} {104026}
  (\bibinfo {year} {2017})},\ \Eprint {http://arxiv.org/abs/1708.07851}
  {arXiv:1708.07851 [gr-qc]} \BibitemShut {NoStop}%
\bibitem [{\citenamefont {Blanchet}(2014)}]{Blanchet:2013haa}%
  \BibitemOpen
  \bibfield  {author} {\bibinfo {author} {\bibfnamefont {Luc}\ \bibnamefont
  {Blanchet}},\ }\bibfield  {title} {\enquote {\bibinfo {title} {{Gravitational
  Radiation from Post-Newtonian Sources and Inspiralling Compact Binaries}},}\
  }\href {\doibase 10.12942/lrr-2014-2} {\bibfield  {journal} {\bibinfo
  {journal} {Living Rev. Rel.}\ }\textbf {\bibinfo {volume} {17}},\ \bibinfo
  {pages} {2} (\bibinfo {year} {2014})},\ \Eprint
  {http://arxiv.org/abs/1310.1528} {arXiv:1310.1528 [gr-qc]} \BibitemShut
  {NoStop}%
\end{thebibliography}%
\end{document}